\documentclass[12pt,preprint]{aastex}

\slugcomment{To appear in ApJ}

\shorttitle{Substructure in the Halo}
\shortauthors{Keller et al.}

\begin{document}


\title{Revealing Substructure in the Galactic Halo - \\ The SEKBO RR Lyrae Survey}


\author{Stefan C.\ Keller, Simon Murphy, Sayuri Prior, Gary DaCosta, and Brian Schmidt}
\affil{Research School of Astronomy and Astrophysics, Australian National University, Canberra, Australia}


\begin{abstract}
We present a search for RR Lyrae variable stars from archival observations of
the Southern Edgeworth-Kuiper Belt Object survey. The survey covers 1675
square degrees along the ecliptic to a mean depth of $V$=19.5, i.e. a
heliocentric distance of $\sim50$kpc for RR Lyrae stars. The survey reveals
2016 RR Lyrae candidates. Follow-up photometric monitoring of a subset of
these candidates shows $\sim$24\% contamination by non-RR Lyrae variables. We
derive a map of over-density of RR Lyraes in the halo that reveals a series of
structures coincident with the leading and trailing arms of debris from the
Sagittarius dwarf galaxy. One of the regions of over-density is found on the
trailing arm, 200$^{\circ}$ from the main body of the Sagittarius dwarf at a
distance of $\sim45$kpc. This distant detection of the stellar population of
the outer trailing arm of Sagittarius offers a tight constraint on the motion
of the dwarf galaxy. A distinctly separate region of over-density is seen
towards the Virgo Over Density.
\end{abstract}





\section{Introduction}

The stellar halo of our galaxy offers a window to the past. Stars in the halo have long orbital periods and inhabit regions of space where the galactic potential is relatively smooth and slowly evolving. Consequently, structures in the halo suffer reduced spatial and kinematic dissipation and may persist for several Gigayears \citep{Bullock05}. Such halo structures offer an insight into the process of galaxy formation, arguably one of the most important processes yet to be understood by modern astrophysics.

One schematic for the process of halo formation is that due to \citet{ELS}, who found that the stars on the most eccentric orbits were also the most metal-poor. This was interpreted as the signature of monolithic collapse of the proto-galactic cloud; the most metal-poor, and hence oldest, stars forming during the collapse. This solitary process of galaxy formation was challenged by \citet{SZ}, who proposed the large dispersion seen in the metallicity of globular clusters and their lack of radial abundance gradient indicated that the Galaxy was the composite of a large number of sub-galaxy sized pieces.

Recent observational studies have converged on a combination of scenarios \citep[for a review]{Freeman02}. This is perhaps most clearly demonstrated by the works of \citet{Kinman07}, \citet{Carollo07} and \citet{Miceli07} who find a distinct outer halo population. \citet{Carollo07} shows that whereas the inner halo (galactocentric radius less than 15kpc) is dominated by highly eccentric, prograde orbits and a metallicity of [Fe/H]$\sim-1.6$, the outer halo exterior to this possesses a more uniform distribution of eccentricities, includes highly retrograde orbits and [Fe/H]$\sim-2.2$. \citet{Kinman07} study of the local halo similarly identifies a component that possesses retrograde rotation and streaming motions and another that exhibits negligible rotation and less signs of streaming. Kinman et al.\ furthermore show that the horizontal branch morphologies of the two components of the halo are consistent with those seen in the young and old globular clusters, reinforcing the conclusions of \citet{Lee99} that the inner halo is consistent with rapid collapse and the outer halo with the ongoing accretion of dwarf galaxies. 

A halo that formed from \emph{in situ} star formation would possess, at the current epoch many dynamical times later, very little substructure. Simulations of galaxy formation in the context of $\Lambda$CDM predict that the majority of the stars in the outer halo formed in progenitor dwarf galaxies and were subsequently accreted \citep{Abadi06,Bullock05}, with the general conclusion that significant amounts of substructure should survive to this time. However, we must point out that $\Lambda$CDM predictions on galaxy scales are contentious at this stage. \citet{Bell07} compare the level of structure evident in the SDSS data with simulations and finds that the Milky Way's halo is well matched to a halo built exclusively from disrupted satellite galaxies.

Numerous studies of the spatial distribution of stars in the halo have shown the importance of ongoing accretion of satellite galaxies on to the Milky Way. The most striking example of ongoing accretion is the Sagittarius dwarf galaxy \citep[Sgr]{Ibata94}. The ensuing trail of debris has been found to wrap around the sky \citep{fieldofstreams,Majewski03,Ibata03,Yanny03,Newberg02}. The debris of Sgr is arguably the largest contributor yet found to the halo substructure at radii of over 50kpc. The orbit of Sgr, as traced by its extensive debris tails, has been used as a probe of the mass and shape of the Galaxy \citep{Ibata01, Helmi01, Helmi04, Johnston05, Law05, Fellhauer06}.

Another halo structure, the Virgo Stellar Stream (VSS) was found as an over-density of RR Lyrae stars (RRLs) \citep{VivasZinn06,Zinn04} at galactocentric radius $\sim20$kpc. Later radial velocity measurements \citep{Duffau06} revealed the grouping shared a common space velocity. \citet{Newberg02}, \citet{Juric05} and more recently \citet{Newberg07}, find a coincident diffuse over-density of F-type main sequence stars spanning some 1000 square degrees of sky called the Virgo Overdensity (VOD). \citet{Newberg07} proposes that the VSS and VOD are the same structure. On the basis of radial velocity measurements Newberg et al.\ concludes it is not associated with a confluence of the leading and trailing arms of Sgr as proposed by \citet{MartinezDelgado07} but is possibly remnant material from another merger event.

The Monoceros Stream \citep{Newberg02} is believed to be a low surface brightness ring encircling the Galaxy at 15-20kpc \citep{Ibata03}. The structure was first traced by \citet{Newberg02} using halo turnoff stars and subsequently by \citet{Majewski03} using M giants. The stream is not thought to be related to Sgr tidal debris or the VSS/VOD, but rather may result from the disruption of the Canis Major dwarf \citep{MartinezDelgado07,Dinescu06}. \citet{Conn07} reports that interpretation of the Monoceros stream as a warp or flare in the galactic disk does not account for the observed stellar densities.  Possibly related to the Monoceros Stream are the Tri-And over-density \citep{RochaPinto04} at $\sim20$kpc and a yet more distant stellar structure at $\sim28$kpc \citep{Martin07}.

Most recently, \citet{Belokurov07} reports the detection of a significant stellar over-density that covers some 80$^\circ$ of  galactic latitude, dubbed the Hercules-Aquila Cloud. This over-density lies at a distance of between 10-20kpc and straddles the Galactic disk offset by 6-15kpc from the Galactic centre. Spectroscopy of probable cloud members shows that the structure possesses a radial velocity offset of $\sim$180km/s from the halo and thick disk. 

The features discussed above are a small fraction of the number that could be expected from simulations of $\Lambda$CDM hierarchical structure formation \citep{Bullock01,Bullock05}. The search for relic substructure could help illuminate the outstanding 'missing satellite problem' in galaxy formation in  $\Lambda$CMD cosmology \citep{Freeman02}. Here, models predict of order 500 small dark matter halos should surround the Milky Way today, in contrast to the $\sim40$ known dwarf galaxies in the Local Group \citep{Wilkinson07}. If the level of substructure does not match that predicted from a large population of dwarf galaxies this may prompt revisions to model assumptions.

\section{RR Lyrae Variables as Probes of the Halo}

RRL variables have a long history as probes of the Galactic halo (see \citet{Wetterer96} for a historical review). They are stars on the core He-burning horizontal branch that fall within the temperature range of the instability strip. They are easily identified on the basis of their color and variability. In addition, their absolute magnitudes are known to better than 0.1 magnitudes, making them standard candles for the halo. They represent an old stellar population and so by virtue of their age can probe disruption events that occurred at early times. The abundance of RRLs is dependent on the morphology of the horizontal branch of the progenitor population. Local dwarf spheroidals exhibit abundant RRL populations, hence these variable stars should be good tracers of accretion events.

In recent years several sky surveys have discovered large numbers of RR Lyrae variables. The survey area, limits and completeness of these surveys are summarised in Table \ref{table:surveys}.  The QUEST RRL survey  \citep{VivasZinn06} proposes six significant substructures; three associated with the Monoceros Stream, and others associated with Pal5, VSS, and the Sagittarius Stream. Searches for RRLs in the SDSS are presented by \citet{Sesar07}, \citet{Ivezic04} and \citet{Ivezic00} from multiple imaging and by \citet{Ivezic05} from colors alone. \citet{Sesar07} reviews the SDSS work to date, proposing 13 new over densities. 

The QUEST and SDSS RRL surveys cover equatorial regions. The LONEOS-1 Survey \citep{Miceli07} traces the ecliptic but to a moderate depth (reaching distances of up to 30kpc). With between 28-50 epochs per star \citet{Miceli07} are able to identify the Oosterhoff type of each RRL. They conclude that the radial distribution of Oosterhoff type I and II (OoI and OoII) RRLs are systematically different. \citet{Lee99} found globular clusters that are dominated by OoI RRLs show little net angular momentum as expected from a formation scenario such as \citet{SZ} whereas OoII globular clusters show net prograde rotation as would result from the \citet{ELS} scenario. \citet{Miceli07} propose that the presence of distinct OoI and OoII populations reveals that both the monolithic collapse and accretion scenarios are important for constructing the halo. This conclusion parallels the two kinematic components found in the halo by \citet{Carollo07}. The presence of substructure in their sample of RRLs is not addressed in \citet{Miceli07}.

\section{The Southern Edge\-worth\--Kuiper Belt Object Survey}

The present study is based upon the archived observations of the Southern Edgeworth-Kuiper Belt Object (SEKBO) survey \citep{Moody03}. SEKBO was a multi-epoch, two color imaging survey of a 10 degree band centred on the ecliptic. Observations were obtained over the three years from December 1999 on the Mount Stromlo Observatory 50-inch Great Melbourne Telescope that was previously used for the MACHO microlensing survey. The telescope featured a dual-band imaging system in which a dichroic allowed for simultaneous exposures in two broad bands on separate 4k$\times$4k pixel CCD mosaics. The MACHO blue filter, hereafter $B_{\rm{M}}$, spans 455-590nm (wavelengths at 50\% normalised response) and the red, $R_{\rm{M}}$, from 615-775nm \citep{BessellGermany99}. These wide, non-standard filters most closely match Johnson $V$ and $R$. 

\begin{figure}[h]
\begin{center}
\includegraphics[scale=0.36, angle=0]{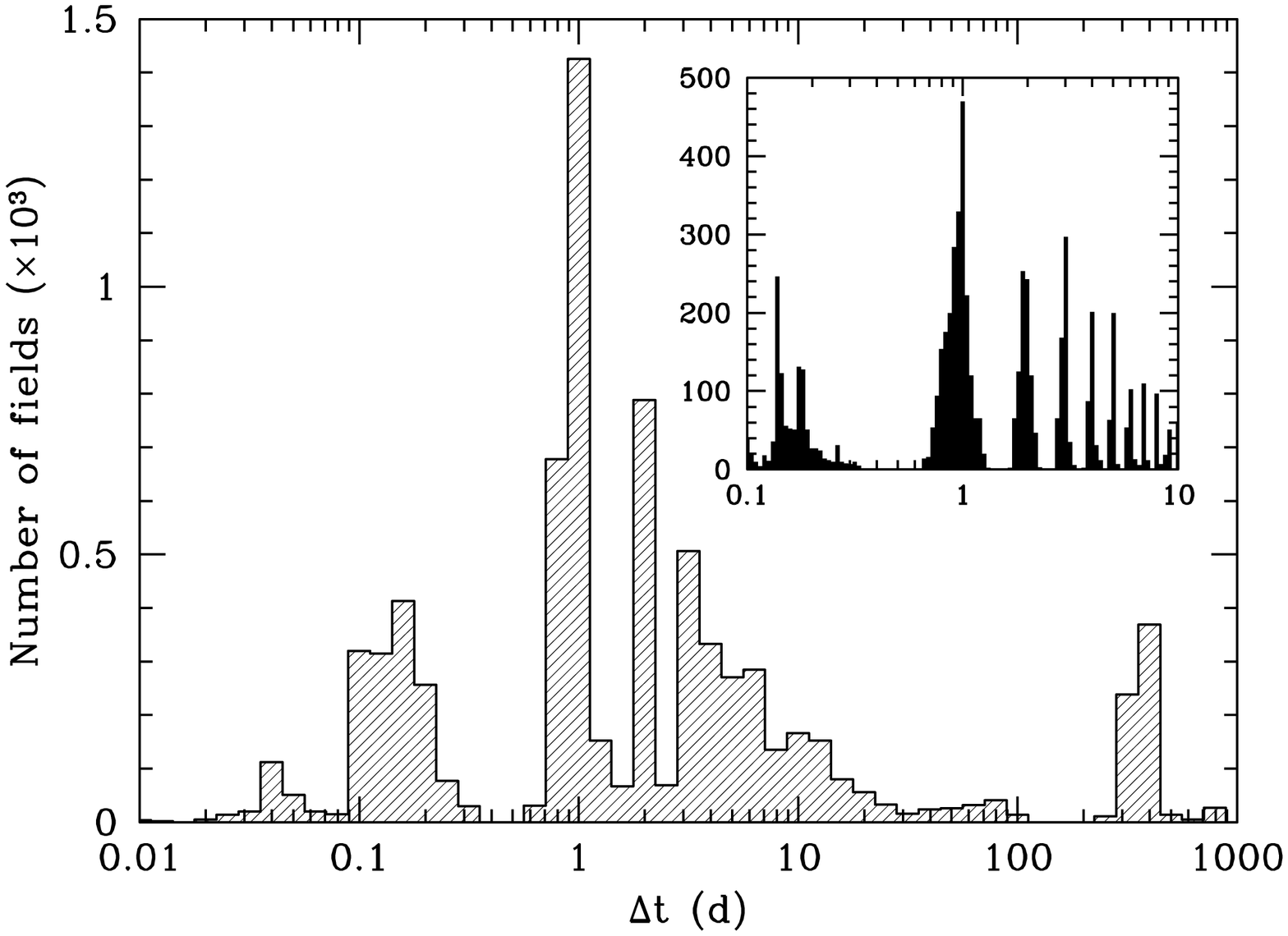}
\caption{The time difference in days between epochs for the SEKBO fields. As expected from the survey design there are clusters of observations at 1-4 hours and 1-7 days apart. The group at 1 year are repeat observations of fields that failed the previous year. The insert shows a the distribution between 0.1-10d.}\label{figure:deltat}
\end{center}
\end{figure}

\begin{figure}[h]
\begin{center}
\includegraphics[scale=0.46, angle=0]{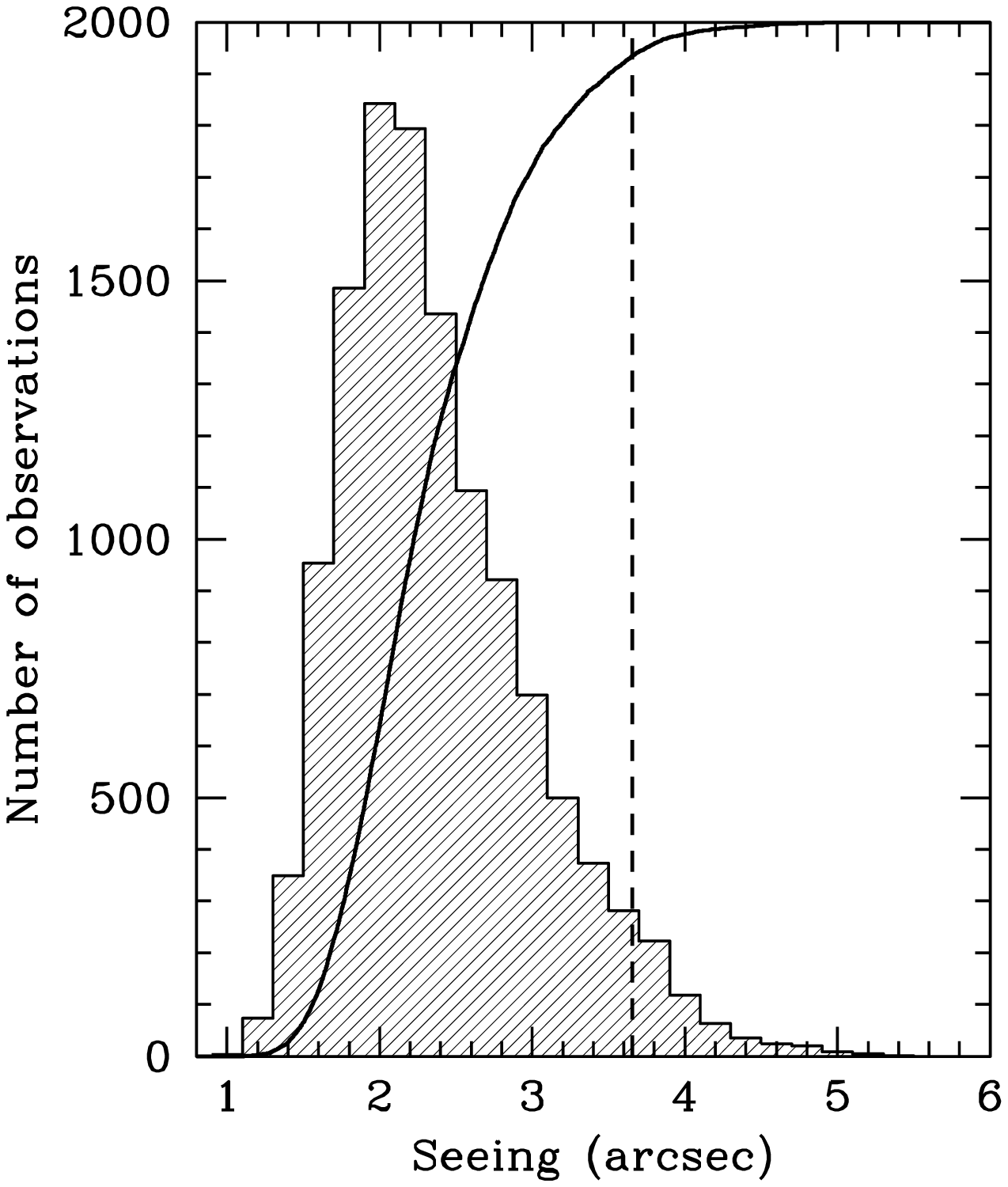}
\caption{The seeing distribution of images in the present study. The solid line is the cumulative distribution, while the dashed line shows the 3.6" seeing cut imposed on the observations as detailed in \S\ref{section:stdphot}}\label{figure:seeing}
\end{center}
\end{figure}

SEKBO was designed to search for Trans-Neptunian Objects from a set of three exposures of a given field, the first and second epochs separated by $\sim$4 hours and a subsequent third epoch 1-7 days later. As the telescope operated in an automated manner with limited environmental monitoring some observations were obtained under conditions of insufficient quality. These would break an observing sequence for a field and require reacquisition of the field at a later date. Figure \ref{figure:deltat} shows the distribution of time intervals between repeat observations of the fields in our sample. The exposure time was 300 seconds. Under reasonable photometric conditions, this results in a useful magnitude range for photometry of $14<V<20$. Figure \ref{figure:seeing} shows the distribution of seeing derived from the SEKBO images. The seeing is typically 2" with a tail out to $>5$".

\begin{figure}[h]
\begin{center}
\includegraphics[scale=0.46, angle=0]{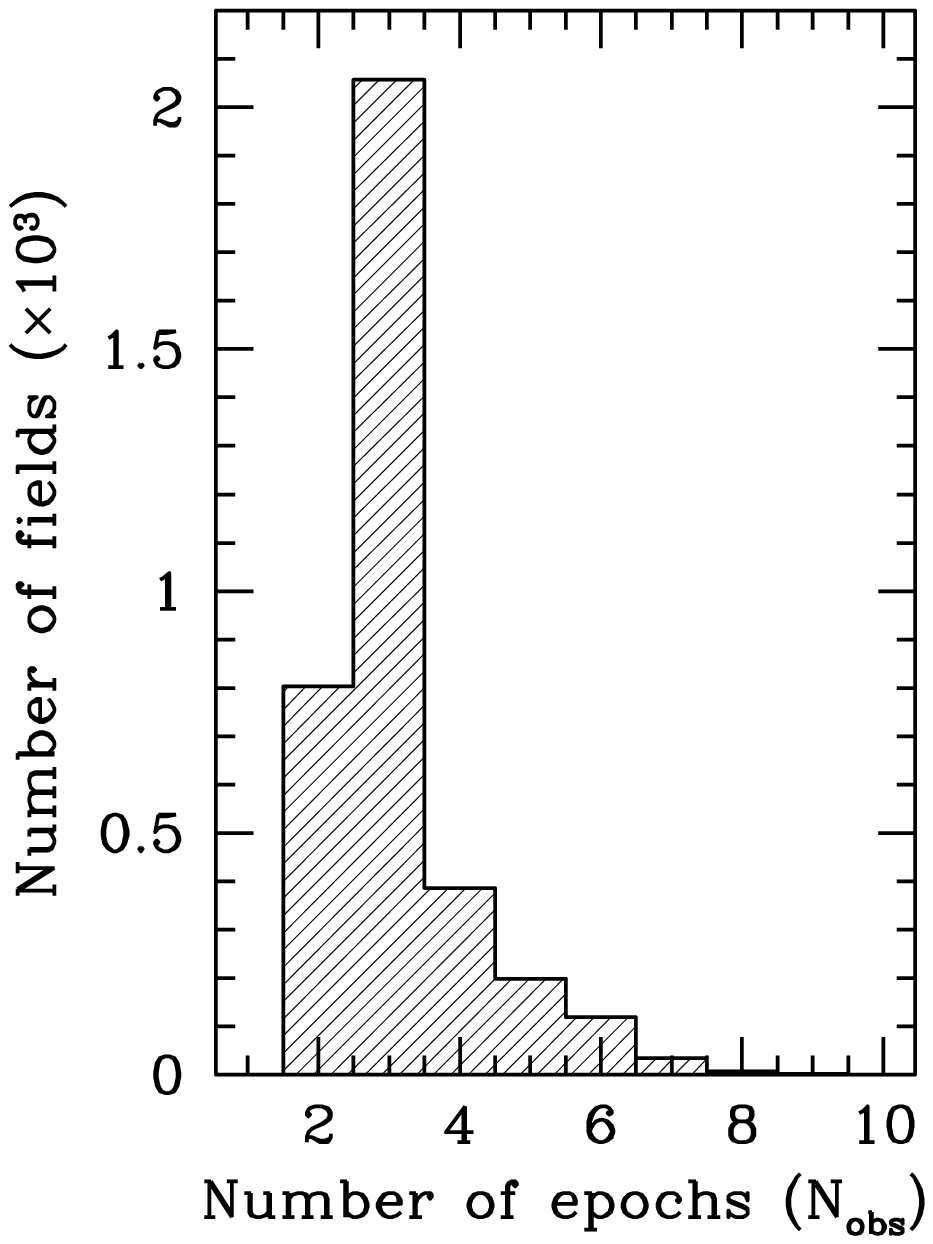}
\caption{Histogram of the number of images per field for the sample of 5212 fields.}\label{figure:numimages}
\end{center}
\end{figure}

\begin{figure*}[h]
\begin{center}
\includegraphics[scale=0.75, angle=0]{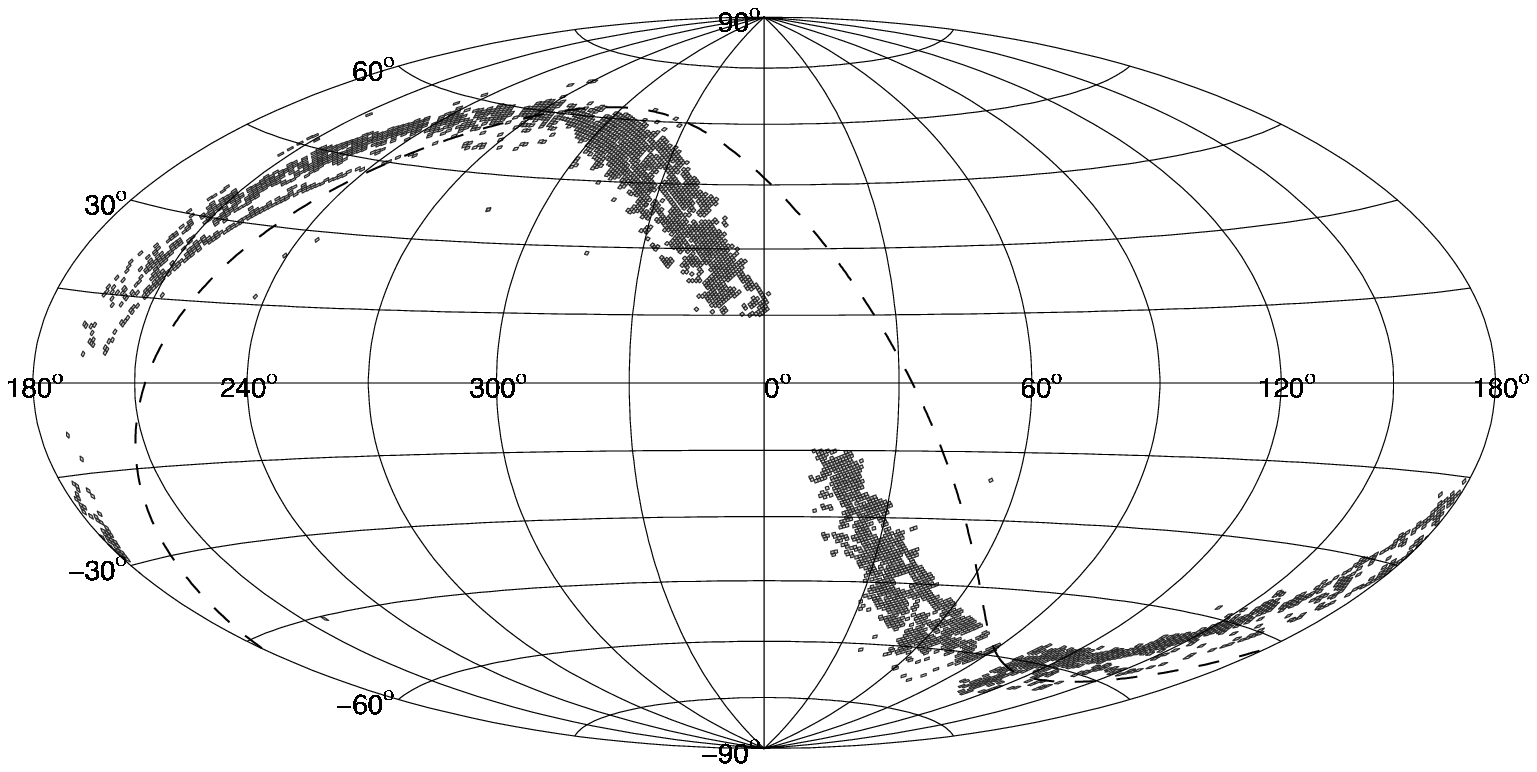}
\caption{The distribution of fields in the present survey in galactic coordinates. The dashed line is the celestial equator.}\label{figure:fields}
\end{center}
\end{figure*}

There are 5212 field centres with two or more epochs that form the basis of our sample. The histogram of the number of epochs per field in the sample is shown in Figure \ref{figure:numimages}. Most fields have 2 or 3 observations but there is a tail out to 10 observations. Figure \ref{figure:fields} shows the sky coverage of the survey. Note that we have excluded regions with $E(B-V)>0.10$ as well as  galactic latitudes $b<10^{\circ}$ to avoid fields with excessive crowding. The total effective area of the survey (having corrected for inter-CCD gaps and bad columns, see S\ref{section:completeness}) is 1675 square degrees.

\section{Photometric Reduction}

In this section we outline the steps taken to reduce the raw frames to calibrated photometry. Each of the 
17578 exposures considered consists of 16 images (2 colors $\times$ 4 CCDs $\times$ 2 amplifiers) which are stored on the Australian National University's Supercomputing facility mass data store. The data set is 2.3Tb in size.

The data in the archive has been processed to remove bias and flatfield structure using the data reduction pipeline of the MACHO system. This system maintained a moving average bias and flatfield that was regularly checked for changes in the optical/detector system \citep{Alcock99}.

We first determine a world co-ordinate system (WCS) for each image based upon the nominal pointing of the telescope as recorded in the image FITS header. UCAC2 stars \citep{UCAC2} in common to the image are used to derive the WCS. The UCAC2 provides between 40 and several hundred stars per CCD. Our derived positional uncertainties are better than 50 milliarcsecs for bright stars. A ZPN\footnote{Zenithal/azimuthal polynomial projection see \cite{Calabretta04}} coordinate representation is used. This representation utilizes a radial term to characterise distortions across the image and a simple linear transformation to remove the effects of rotation, scale, and any shear caused by refraction.  Linearising the telescope system in this way provides robust coordinate transformations across multiple images of a field with no difficulties in CCD corners or in images with few astrometric standards.

Differences in the fabrication and electronics associated with each CCD/amplifier pair result in a non-uniform gain across the red and blue mosaic. Gain matching between each CCDs two amplifiers was achieved by matching the median sky background level. The four CCDs were then mosaiced together using the {\tt{SWarp}} package\footnote{{\tt{http://terapix.iap.fr/}}, \citet{Bertin05}}. The median sky brightness was then used to scale the four CCDs of each passband to the same gain level.

\subsection{Photometry and Object Classification}

We base our object catalogue upon a template image that is created for each field by co-adding the individual $R_{\rm{M}}$ observations. In the case of three or more images we use {\tt{SWarp}} to form a median image to robustly reject bad pixels and cosmic-rays. In the case of two images for a field we apply an algorithm that examines corresponding pairs of pixels in the registered image stack and removes high values from pairs with large variance. The footprint of the template image is only that area on the sky that is represented in all images of the field.

The template image is used solely for object detection. The combination of images taken under different seeing would render the extraction of photometry from the template problematic due to the distorted and non-photometric PSF shape that results from combination. The template image does have several advantages for object detection, however, including greater depth and reduced numbers of false objects due to the exclusion of cosmic-rays and bad pixels. Object detection within the template image was made using the {\tt{cdoPhot}}\footnote{{\tt{http://www.vuw.ac.nz/staff/michael\_reid/software.html}}} package.

Photometry was then performed on the individual images that comprised the template. This photometry was performed in three modes: in fixed position aperture mode where the template coordinates are used for the centre of the aperture; centroid aperture mode where the barycentre of the object is used for the centre of the aperture; and PSF-fitting mode where a PSF is constructed for each object. Aperture corrections were applied to correct the magnitude defined through a 3 pixel radius aperture to the aperture used by \citet{Landolt92} for standard star photometry (discussed below). 

The PSF-fitting mode is the basis of our object classification. The gaussian widths in the x and y directions give a clear discriminant as to the nature of the object. Stellar objects lie clustered around a particular x-width, y-width. Galaxies exhibit more extended dimensions and cosmic-rays a smaller extent. To make a quantitative judgement as to the stellar nature of the objects in our data we construct a \emph{stellarity index} that expresses the distance an object lies in gaussian x-width, y-width space from the stellar locus.

\subsection{Standardised Instrumental Photometry}\label{section:stdphot}

Raw magnitudes from the gain matched frames were then placed on a standard instrumental system. Due to non-photometric conditions and atmospheric extinction there were zero-point offsets between magnitudes in each observation and that obtained from the detector system at the zenith under photometric conditions (hereafter the standard instrumental system). The 2MASS \citep{Skrutskie06} $J$ and $K$ magnitudes allow this offset to be calculated. From images obtained in photometric conditions we first derived a linear fit:
\begin{equation}\label{eqn:1}
R_{\rm{M}, i}-J = -2.5 \log f_{R_{\rm{M}, i}} + 2.5 \log t_{exp} -J\\
= \alpha(J-K) + ZP_{R_{\rm{M}, i}}
\end{equation}
where $f_{R_{\rm{M}, i}}$ is the flux in MACHO $R$ for the $i$th observation, $t_{exp}$ = 300 seconds and $ZP_{R_{\rm{M}, i}}$ is a zeropoint defined below. We determine $\alpha=1.100$ over a narrow color range ($0.3<J-K<0.6$).

For each individual image in the data set, a fit was then performed on the order of 100 stars in each field with signal-to-noise ratio $>$ 20  ($J<15$ and $K<14.5$) within the color range above. We define our standard instrumental system such that a $R_{\rm{M}}=22$ star provides 1 ADU s$^{-1}$. Hence:
\begin{equation}\label{eqn:2}
R_{\rm{M}, i}-ZP_{R_{\rm{M}, i}} = 22
\end{equation}
where $ZP_{R_{\rm{M}, i}}$ is the zeropoint required to draw the observation onto the standard instrumental system ($ZP_{R_{\rm{M}, i}}$ is the extrapolation of the linear relation of Equation \ref{eqn:1} to $J$$-$$K$=0).  

The effective wavelength of the $B_{\rm{M}}$ filter is sufficiently blue of the 2MASS bands that the above procedure can not be used as it would introduce a significant interstellar extinction dependance.  To determine the $B_{\rm{M}}$ zeropoint we used the $V$$-$$R$ color of sources from \citet{Stetson00} and their corresponding $B_{\rm{M}}$$-$$R_{\rm{M}}$ colors. Note that because the blue and red exposures are obtained simultaneously there is a single offset between $B_{\rm{M}}$ and $R_{\rm{M}}$ (with a minor airmass dependence, see below) and this is found by the requirement that 
\begin{equation}\label{eqn:3}
V-R = 0 =(B_{\rm{M}}-ZP_{B_{\rm{M}}}) - (R_{\rm{M}}-ZP_{R_{\rm{M}}})
\end{equation}

Since the MACHO filters are well matched to the standard $V$ and $R$ only a linear regression is required. However, due to the chromatic effect of atmospheric extinction, the difference in zeropoints will be a function of airmass, $X$. From the observed Landolt fields we derive,
\begin{equation}\label{eqn:4}
ZP_{B_{\rm{M}}} - ZP_{R_{\rm{M}}} = (0.316\pm0.06) - (0.08\pm0.02)(X-1)
\end{equation}
and scale the flux of the blue images to provide 1 ADU s$^{-1}$ for a $B_{\rm{M}}=22$ star.

We reject those observations displaying zeropoint magnitudes that show more than 0.75 magnitudes of extinction or sky brightness that are more than 1 magnitude brighter than the median. We also remove those observations with seeing in excess of 3.6" to reduce the problems introduced into the RRL detection technique by objects that are blended in poorer seeing frames and resolved in nominal seeing frames.This leaves 3692 field centres with acceptable data.

\subsection{Transformation to $V$ and $R$}

A number of fields in the dataset contain standard stars observed by \citet{Stetson00}. This allows us to define the transformation equations to take our standard instrumental photometric system to that of Landolt $V$ and $R$ \citep{Landolt92}. Given the close matching of the MACHO bandpasses to the $V$ and $R$ filters only a linear transformation in color is required. The transformation is of the form:
\begin{equation}\label{eqn:5}
V = B_{\rm{M}} + \alpha\ (B_{\rm{M}}-R_{\rm{M}}) + \beta
\end{equation}
for constants $\alpha$ and $\beta$, and where $B_{\rm{M}}$ and $R_{\rm{M}}$ have been both zero--point offset and aperture corrected. Observations of a large sample of standard stars allow us to define the following conversions to Landolt $V$ and $R$ magnitudes
\begin{equation}\label{eqn:6}
V = B_{\rm{M}} - 0.12\ (B_{\rm{M}}-R_{\rm{M}}) + 0.19
\end{equation}
and,
\begin{equation}\label{eqn:7}
R = R_{\rm{M}} + 0.15\ (B_{\rm{M}}-R_{\rm{M}}) + 0.34
\end{equation}

\begin{figure}
\begin{center}
\includegraphics[scale=0.38, angle=0]{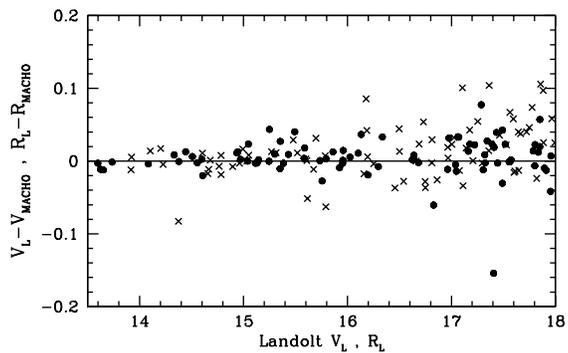}
\caption{The difference between the Landolt $V$ (circles) and $R$ (crosses) standard magnitudes and those derived from our photometry for the \citet{Stetson00} stars in our sample.}\label{figure:landolt}
\end{center}
\end{figure}

Figure \ref{figure:landolt} shows the difference between the magnitudes in $V$ and $R$ derived from Equation \ref{eqn:6} and those of \citet{Stetson00} from the 160 stars in common. This figure shows an offset of $\Delta V$ = 0.01 $\pm$ 0.03 and $\Delta R$ = 0.01 $\pm$ 0.03. It should be noted that the photometry that contributes to Figure \ref{figure:landolt} was obtained under hetrogeneous  seeing, atmospheric extinction and photometric conditions over the course of three years giving us confidence in our global zeropoints to the level of $\pm$0.03 magnitudes.

\section{RR Lyrae Candidate Selection}
\label{section:candidates}
We have implemented a probabalistic approach to the selection of RRL candidates. We first remove those objects with shape and color parameters grossly unlike the stellar population. Selection is then made based on color and variability. In order to quantify the significance of variation we need to first understand the photometric uncertainties.

\begin{figure}
\begin{center}
\includegraphics[scale=0.38, angle=0]{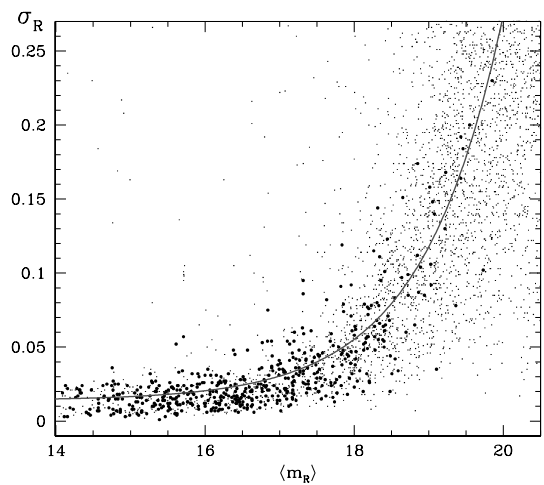}
\caption{The standard deviation of the photometry in $R_{\rm{M}}$ as a function of magnitude for typical field in our survey.  Overlaid is the sky-limited Poisson curve (line: see text) that is a fit to objects with unambiguously stellar shape parameters (the bold points).}\label{figure:photuncertainty}
\end{center}
\end{figure}

Photometric uncertainties were charactersised using the observed standard deviations of the stellar objects from the multiple epochs. Figure \ref{figure:photuncertainty} shows the distribution of standard deviation versus magnitude for objects in a particular survey field. For most stars the standard deviations are indicative of the photometric uncertainties for the observation. However, variables stars or objects with bad photometry on some epochs will lie above the main locus.

To determine the mean photometric uncertainty as a function of magnitude we fit a sky-limited Poisson distribution:
\begin{equation}\label{eqn:sigma}
\sigma_{phot} = 10^{+0.4(m+m_{cut})}+m_{floor}
\end{equation}
for magnitude $m$, where $m_{cut}$ is iteratively determined so that the standard deviation of half the objects lie above the fitted line and half below. The constant $m_{floor}$ is the component of the photometric uncertainty that is not related to the photon statistics. This is derived from the brighter stars up to two magnitudes fainter than the saturation limit. The photometric uncertainties vary from field to field. The typical uncertainty for the brighter stars is 0.02 magnitudes rising to 0.2 magnitudes at $V\sim19.5$.

With an understanding of our photometric uncertainties we then construct a RRL index, $\mathcal{L}$:
\begin{equation}\label{eqn:probability}
\mathcal{L} = \mathcal{L}_{Color} + \mathcal{L}_{RRL} - \mathcal{L}_{Not Variable}
\end{equation}
where $\mathcal{L}_{Color}$ quantifies the probability that a star has colors commensurate with an RRL; $\mathcal{L}_{RRL}$ is the log likelihood that the variation in color exhibited by the object matches that expected from RRLs; and $\mathcal{L}_{Not Variable}$ is the log likelihood that the variation can be explained by photometric uncertainty alone. The RRL index, $\mathcal{L}$, is a figure of merit of how much like an RRL each object is. It is not based on a deep statistical foundation, but rather the pragmatic need to distinguish RRLs from other stars with a minimum of observations and with minimal manual intervention. 

We now discuss the above three quantities in more detail. To construct $\mathcal{L}_{Color}$ we first derived the de-reddened color distribution from $\sim19000$ LMC RRLs from the MACHO project (using a mean reddening of E($V$$-$$R$) =0.08 towards the LMC \citep{Alcock04}). This distribution can be approximated by a gaussian with $\langle V-R \rangle_{0}= 0.137$ and $\sigma_{V-R}=0.08$ mag. Given this distribution the probability density of a star having an RRL-like color is:
\begin{equation}\label{eqn:colorprob}
\centering
\mathcal{L}_{Color}(V-R)_{0,i} = \frac{1}{\sigma_{V-R}\sqrt{2\pi}}e^{-\frac{1}{2}( \frac{(V-R)_{0 i}-\langle V-R \rangle_{0}}{\sigma_{V-R}})^{2}}
\end{equation}
where the dereddened $V-R$ color of the $i$-th observation, $(V-R)_{0 i}$, is derived from Equation \ref{eqn:6} and dereddened according to the dust maps of \citet{Schlegel98} (we adopt E($V$$-$$R$)=0.608E($B$$-$$V$) and A$_{V}$=3.315E($B$$-$$V$)).

\begin{figure}
\begin{center}
\includegraphics[scale=0.38, angle=0]{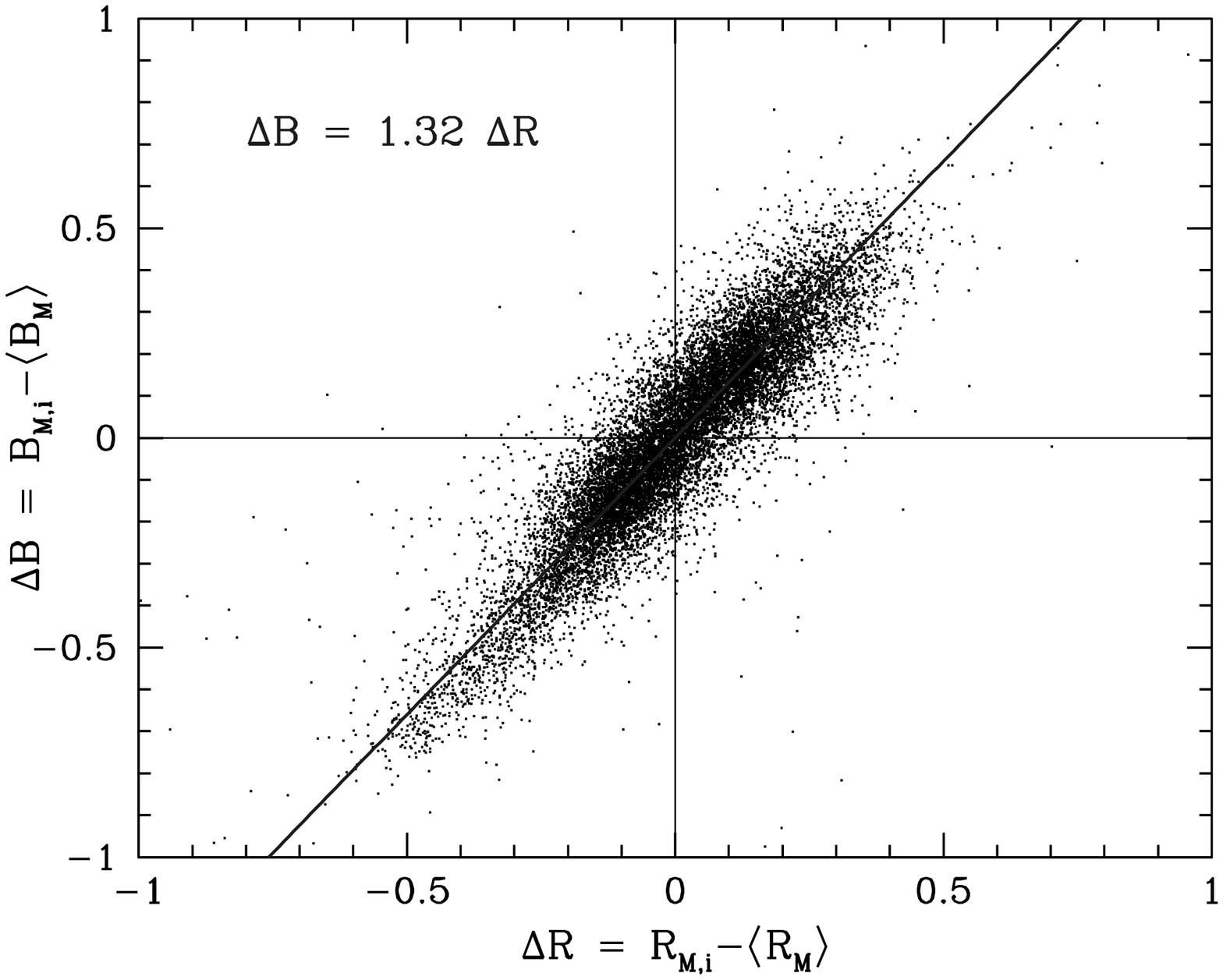}
\caption{The observed color variation of a sample of 119 MACHO RR Lyrae from the LMC. The line shows the best-fit relation $\Delta B = 1.32\Delta R$}\label{figure:RRLcolors}
\end{center}
\end{figure}

\begin{figure}
\begin{center}
\includegraphics[scale=0.44, angle=0]{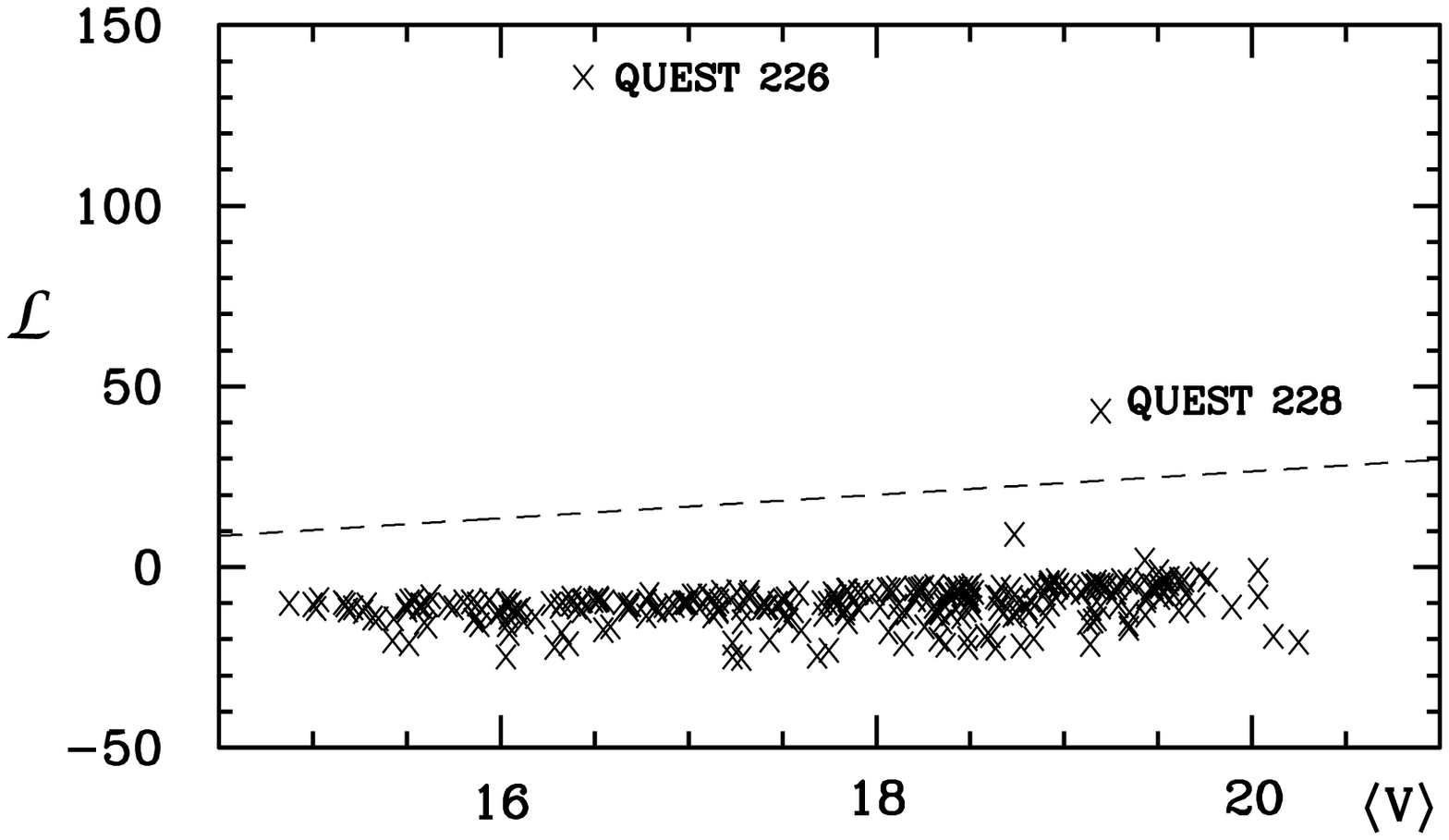}
\caption{RRL index $\mathcal{L}$ versus $V$ magnitude for one of the survey fields that overlaps the QUEST \citep{VivasZinn06} survey. The two QUEST RRLs in the field are recovered above the dashed line that separates the non-RRLs of low $\mathcal{L}$ from the candidates with high $\mathcal{L}$.}\label{figure:RRLindex}
\end{center}
\end{figure}

$\mathcal{L}_{RRL}$ compares the observed variation for an object to the variations seen from a sample of RRLs. Densely sampled (N$_{\rm{obs}} \sim 1000$) $B_{\rm{M}}$ and $R_{\rm{M}}$ light curves were obtained for a random sample of MACHO LMC RRLs. For each point in the light curves we calculate $\Delta B_{\rm{M}} = B_{\rm{M}, i} - \langle B_{\rm{M}} \rangle$ and $\Delta R_{\rm{M}} = R_{\rm{M}, i} - \langle R_{\rm{M}} \rangle$. These are plotted in Figure \ref{figure:RRLcolors}. The line $\Delta B_{\rm{M}} = 1.32\Delta R_{\rm{M}}$ shows a classic signature of pulsation, namely, that the amplitude of pulsation increases with decreasing wavelength. To compare these points to the observed variation in our objects we determine the fraction of MACHO points enclosed within an error ellipse centred on ($\Delta B_{\rm{M}}$,$\Delta R_{\rm{M}}$) in Figure \ref{figure:RRLcolors} for each epoch.
\
$\mathcal{L}_{NotVar}$ measures the probability that a non-variable star of given magnitudes $B_{\rm{M}}$, $R_{\rm{M}}$ and photometric uncertainties $\sigma_{B}$, $\sigma_{R}$ could have the observed variation. $\mathcal{L}_{NotVar}$ for the $i$-th observation is expressed as follows:
\begin{equation}\label{eqn:notvarprob}
\centering
\mathcal{L}_{NotVar, i} = \frac{1}{2\pi\sigma_{B}\sigma_{R}}e^{-\frac{1}{2}\big( \frac{R_{{\rm{M}},i}-\langle R_{\rm{M}} \rangle}{\sigma_{R}}\big)^{2}}e^{-\frac{1}{2}\big( \frac{B_{{\rm{M}},i}-\langle B_{\rm{M}} \rangle}{\sigma_{B}}\big)^{2}}
\end{equation}
The combined RRL index, $\mathcal{L}$, for a typical field is shown in Figure \ref{figure:RRLindex}. Of the 3928 objects identified in this field only 523 objects pass the shape and color restrictions and remain in this figure. The bulk of stars without RRL-like variability and color have indices less than zero. Two objects in the field in question show high indices. We identify RRL candidates above the dashed line in Figure \ref{figure:RRLindex}. This line is derived from the distribution of 20 QUEST RRLs in our survey area (all are RR{\emph{ab}}s). Of these RRLs 8 reside in the non-variable locus and 12 present  $\mathcal{L}>40$. The line slopes upwards with $V$ to avoid the flaring of the stellar locus at fainter magnitudes. Candidate RRLs were then visually inspected to remove objects associated with obvious image flaws (ghosting, bad columns, proximity to bright stars etc.). We also limit candidate colors to $V$$-$$R < 0.3$ as discussed in \S7. This resulted in 2016 RRL candidates.

\section{Survey Completeness}\label{section:completeness}
The survey completeness is the fraction of RRLs recovered by our detection technique from those available in the field. A detailed characterisation of the completeness is necessary for the analysis to follow. The completeness depends critically on magnitude since our ability to recover RRLs decreases as we delve fainter due to increasing photometric uncertainty. In addition, the limiting magnitude varies across our surveyed fields. Therefore it was necessary to calculate the completeness and a function of magnitude for each field.

To measure the completeness we created artificial RRL light curves drawn from the period, amplitude and color distributions of the MACHO LMC RRL sample discussed in Section \ref{section:candidates}. Since RRL$ab$ and RRL$c$ stars have different period and amplitude distributions we considered the two populations separately. \citet{layden98} template light curves were used to construct the RRL$ab$ light curves and sine curves were used for the RRL$c$ stars. The initial phase of the simulated light curve was chosen at random. Subsequent samples of the light curve were made at the intervals of the observations for the field in question to account for the phase coverage of the field in question. 

To these idealised light curves we added gaussian-distributed  photometric uncertainties as described by Equation \ref{eqn:sigma}. To measure the completeness as a function of magnitude we generated artificial light curves in 0.5 magnitude bins from 14$<R_{\rm{M}}<$21. The artificial data were passed through the selection pipeline and the recovered fraction formed the completeness estimate for that field and magnitude bin. The artificial data however, do not suffer from the imposition of bad columns, occasional dead amplifiers and inter-CCD gaps. To correctly account for these features we determine the effective area of each field. The effective area is the area of the field that has science data from each exposure. On average, this is 89\% of the maximum 0.51deg$^{2}$.

\begin{figure}
\begin{center}
\includegraphics[scale=0.38, angle=0]{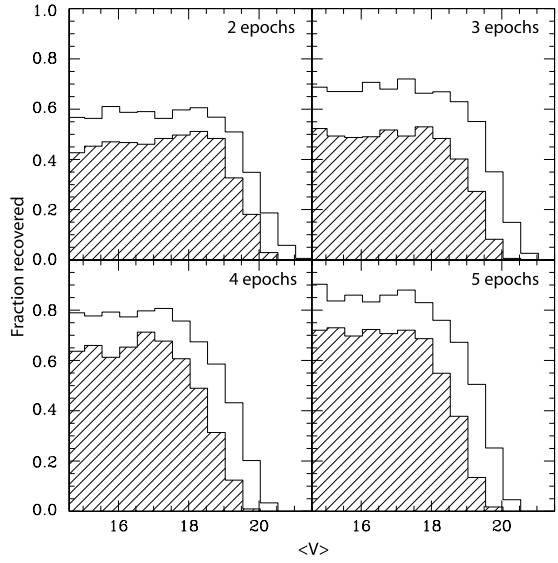}
\caption{Typical completeness profiles as a function of $V$ magnitude for fields with 2, 3, 4, and 5 epochs (a to d respectively). The shaded histograms represent RRL type $c$s. }\label{figure:completeness}
\end{center}
\end{figure}

\begin{figure*}
\begin{center}
\includegraphics[scale=1.1, angle=0]{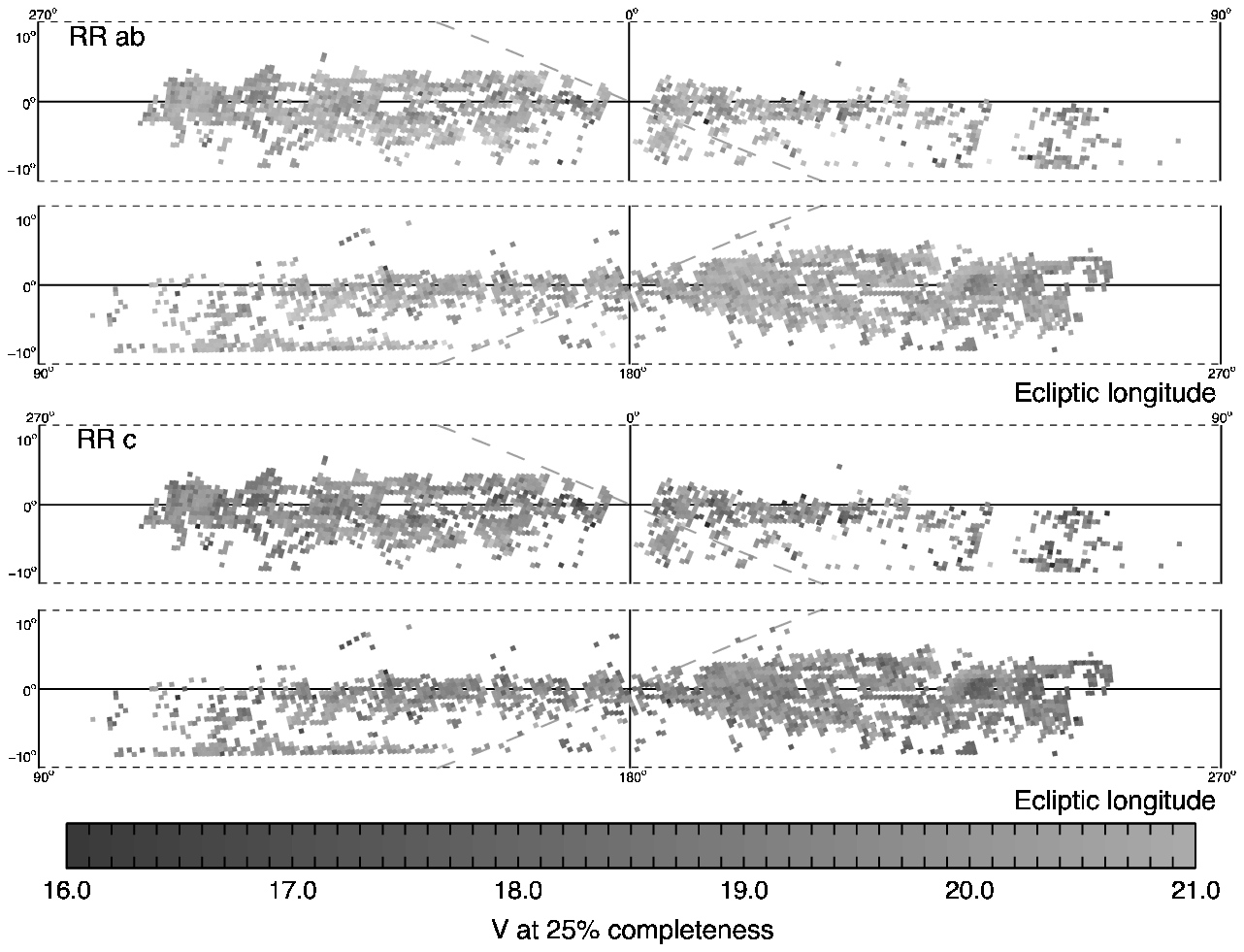}
\caption{A map of completeness across the survey. The color represents the $V$ magnitude at which the completeness drops to 25\% in the case of the RRab (top two panels) and RRc (bottom two panels) variables. The celestial equator is shown as the inclined dashed line.}\label{figure:skycompleteness}
\end{center}
\end{figure*}

Typical completeness profiles are shown in Figure \ref{figure:completeness}. For RRL$ab$s our survey averages $\sim60$\% completeness for $V<18.5$. This falls to a mean completeness of 25\% by $V=19.5$, although a handful of fields extend to $V=19.5$ due to a combination of low sky brightness and exceptionally good seeing. The completeness for RRL$c$s is  systematically lower than that of the RRL$ab$s due to their lower amplitudes. Not surprisingly, the completeness increases with more observations of the field. Figure \ref{figure:skycompleteness} shows a map of the completeness over the survey fields. Fields with anomalous low completeness typically have bad photometry or are derived from observations that are separated by a few minutes and are thus insensitive to the timescale of RRL variation.

The overlap between our survey and QUEST provides a check on our determined completeness. As stated above, of the 20 QUEST RRL$ab$ stars in our survey area, 12 are recovered. This is a magnitude-averaged completeness of 60\%. This compares reassuringly well with the average completeness expected from these fields of 55\%. In the case of RRL$c$ stars the numbers are too small to draw any conclusions; one was recovered out of three expected. 

\subsection{Magnitude Offsets}
The limited number of observations and the asymmetric shape of the RRL$ab$ light curve introduces a bias in the mean magnitude of selected candidates. This can be understood quite simply. Due to the shape of the RRL$ab$ light curve less time is spent near maximum light. Yet an RRL$ab$ has highest likelihood of selection (highest $\mathcal{L}$) when observed at maximum and then again at minimum. This introduces a net bias towards brighter mean magnitudes amongst the RRLs selected as candidates than would be derived from well sampled light curves of the same stars. For bright RRLs the offset is minimal - these objects are detected with high significance from variations that were only a small fraction of the full amplitude due to the low photometric uncertainties. As we proceed to fainter candidates, the fraction of the full amplitude that must be exhibited for the object to stand out as a variable increases as the photometric uncertainty rises.

\begin{figure}
\begin{center}
\includegraphics[scale=0.38, angle=0]{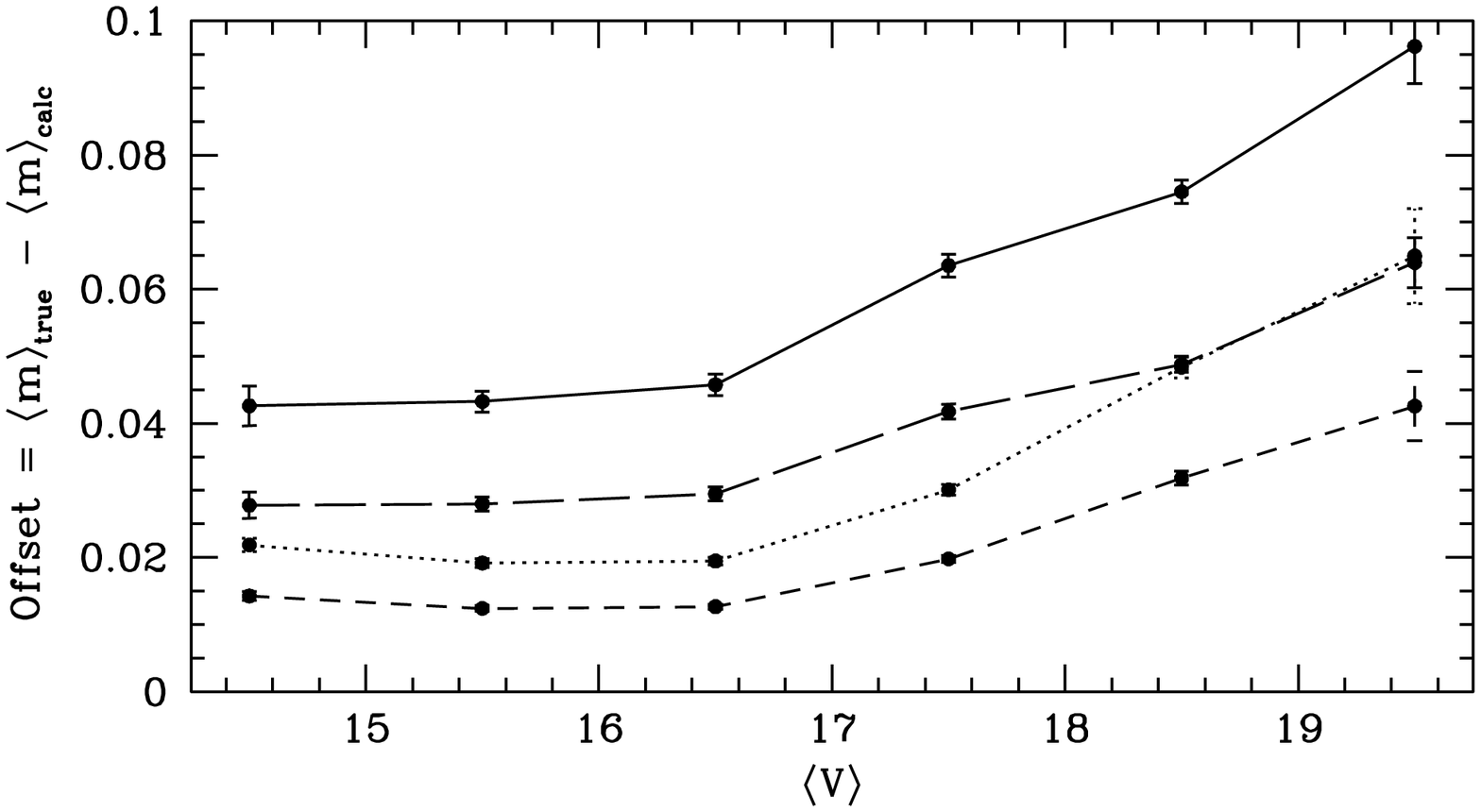}
\caption{The asymmetric nature of the RRL$ab$ light curve introduces a magnitude dependent offset to the mean magnitude derived for our RRL candidates. The mean magnitude offsets for $B_{\rm{M}}$ (the solid line for fields with two observations; the dotted line for more than two observations) and $R_{\rm{M}}$ (the long dashed line for two observations; short dashed for more than two) are shown as a function of magnitude for the ensemble of RRL candidates.}\label{figure:magoffsets}
\end{center}
\end{figure}

Since the photometric uncertainty as a function of magnitude varies from field to field we construct the mean magnitude offset for each individual candidate. A series of artificial RRL$ab$ light curves was generated at the magnitude of the candidate as discussed above. We then determined the offset between the mean magnitude and the mean recovered from the observational epochs. The mean magnitude offset for the ensemble of RRL candidates is shown in Figure \ref{figure:magoffsets}. RRL$c$ stars do not exhibit any significant bias due to their symmetric light curves. Since we do not know apriori whether a candidate is $ab$ or $c$ type we weight the mean magnitude offset by the fraction of each Bailey type to the RRL population. We use a ratio of RR\emph{ab} to RR\emph{c} of 91:9 ratio \citep{Smith95}, although this fraction varies with the Oosterhoff type of the system. For example, in the case of the Draco dSph (OoI) RR\emph{c} contribute 11\% of the 146 known RRLs \citep{Bonanos04} whereas in Boo\"tes dSph (OoII) they comprise 7 of 15 known RRLs \citep{Siegel06}. An OoII halo with 50\% RR\emph{c} would mean that we have systematically underestimated the brightness of each object by at most 0.05 magnitudes towards our faint limit.

\section{Survey Contamination}

\begin{figure}
\begin{center}
\includegraphics[scale=0.38, angle=0]{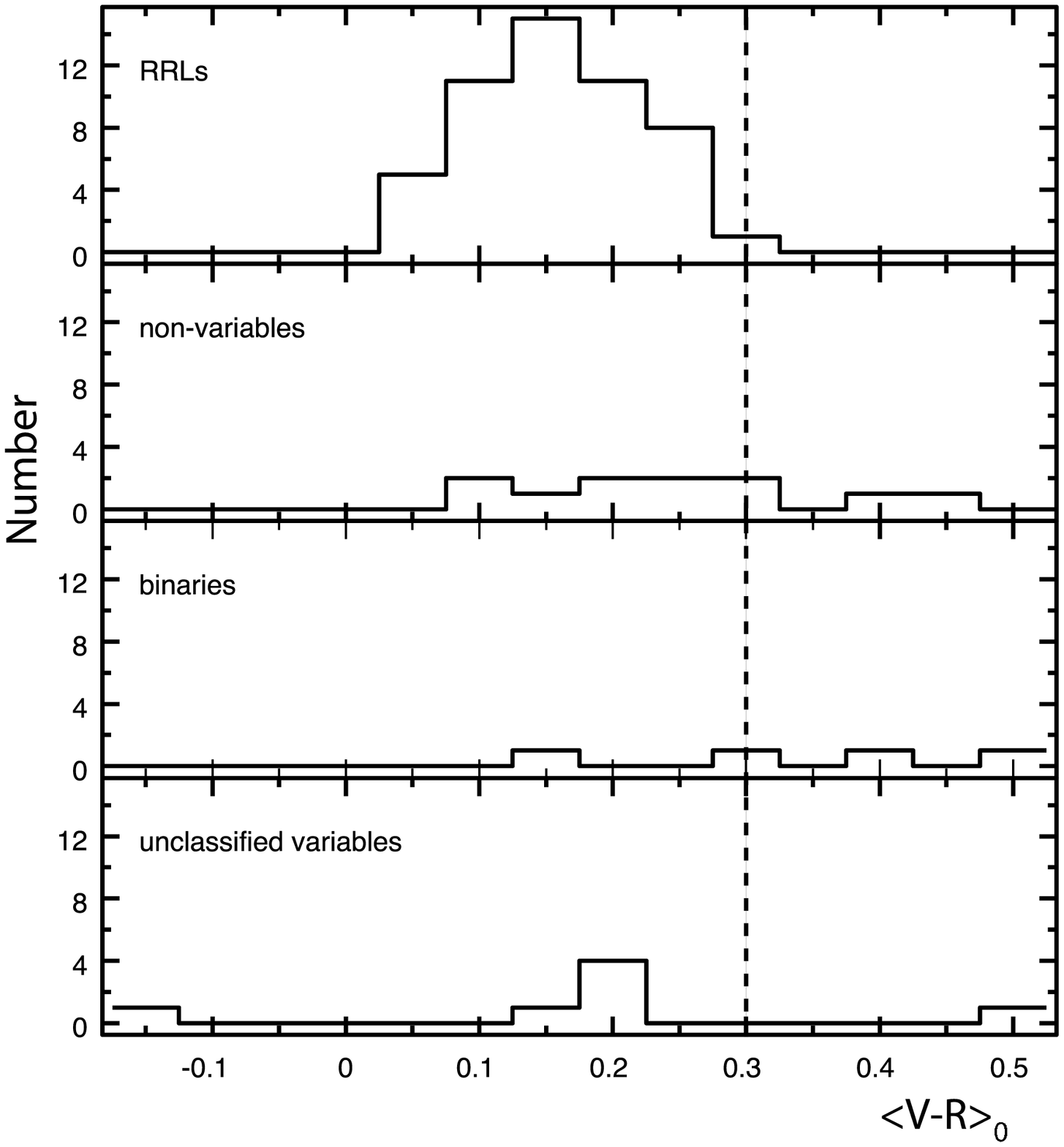}
\caption{The classification of 66 RRL candidates obtained from follow-up photometry.}\label{figure:contamination}
\end{center}
\end{figure}

In order to address the level of contamination by non-RRL objects we performed follow-up observations of a sample of 66 RRL candidates with the Wide-Field Imager on the RSAA 40-inch telescope at Siding Spring Observatory. The selected candidates spanned a range of magnitudes and $V$$-$$R$ color. Light curves of between 8 and 16 points per star were obtained and examined for periodicity (see \citet{Prior07}). The break down of this sample as a function of color is shown in Figure \ref{figure:contamination}. Those that could be phased were either RRLs or eclipsing binaries. A small number of objects were clearly variable yet unclassifiable and a similar number showed no significant variation over the 4-6 nights of observation. 

In the analysis to follow, we apply a color cut to the RRL candidates of $V$$-$$R$$<$0.3. This cut represents a 2 sigma limit of the MACHO LMC RRL color distribution. With such a color cut we find 16 non-RRL amongst 66 candidates (i.e.\ a contamination rate of 24$\pm$12\%).

\section{Properties of the RR Lyrae Candidates}

\begin{figure}
\begin{center}
\includegraphics[scale=0.38, angle=0]{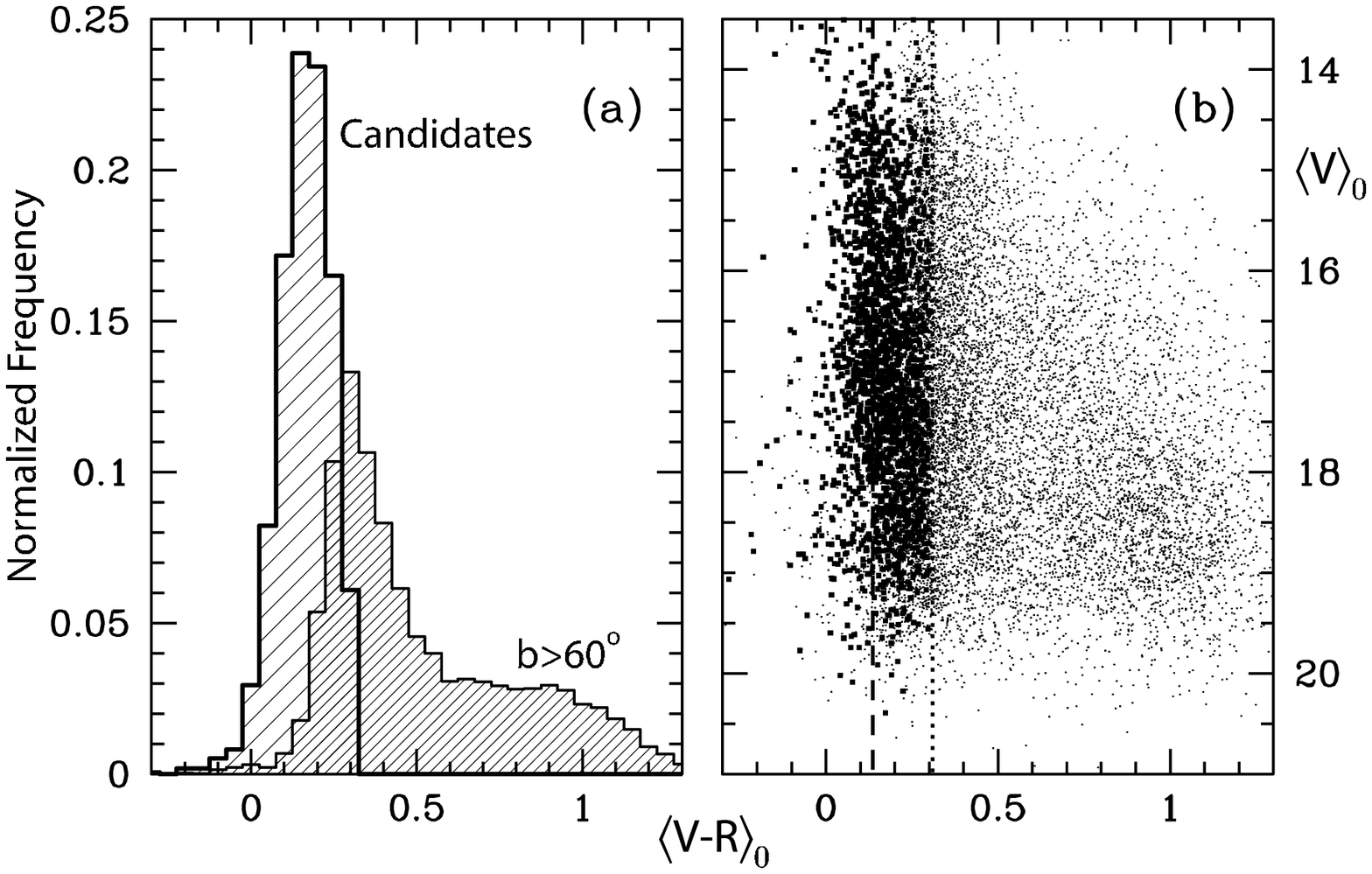}
\caption{(\emph{a}) The normalised distribution of candidate RRL colors compared to that of a representative sample of $\sim$11000 dereddened $b>60^{\circ}$ halo stars. (\emph{b}) the corresponding color-magnitude diagram for the candidate sample and halo population. The long dashed line is the mean $V$$-$$R$ of the MACHO RRL distribution discussed in \S5. The dotted line at $V$$-$$R$=0.3 is the imposed color cut. }\label{figure:candidate_colors}
\end{center}
\end{figure}

The selection process discussed above results in 2016 RRL candidates. Figure \ref{figure:candidate_colors} shows the color-magnitude diagram for the candidate RRLs relative to the general halo background. The halo color-magnitude diagram shows a steep drop off in density for stars of $V$$-$$R<$0.3; this is the color of the halo main sequence turn off. The distribution of candidates is bound to the red by our color limit at $V$$-$$R=$0.3 and to the blue is consistent with the distribution of MACHO LMC RRLs. 

We adopt an absolute magnitude of $M_{V}=0.56$ for the RRL population. This value is based upon the mean absolute magnitude of $M_{V}$=0.59$\pm$0.03 for an [Fe/H]=-1.5 provided by \citet{Cacciari03} who reviewed the absolute magnitude determinations for RRL from a variety of techniques. The mean metallicity of a sample of halo RRLs from \citet{Suntzeff91} is [Fe/H]=-1.65$\pm0.3$ dex, therefore we apply an offset to the value of \citet{Cacciari03} using the metallicity relation of \citet{Chaboyer99}. Our assumed absolute magnitude is equivalent to that used in the study of \citet{VivasZinn06}. Vivas \& Zinn (ibid) find that the effects of evolution away from the zero-age horizontal branch and dispersion in metallicity should lead to a dispersion in $M_{V}$ of 0.13 magnitudes, or in other words a 7\% dispersion in distance.

\subsection{Spatial and Radial Distribution of RR Lyraes}

\begin{figure*}
\begin{center}
\includegraphics[scale=0.55, angle=0]{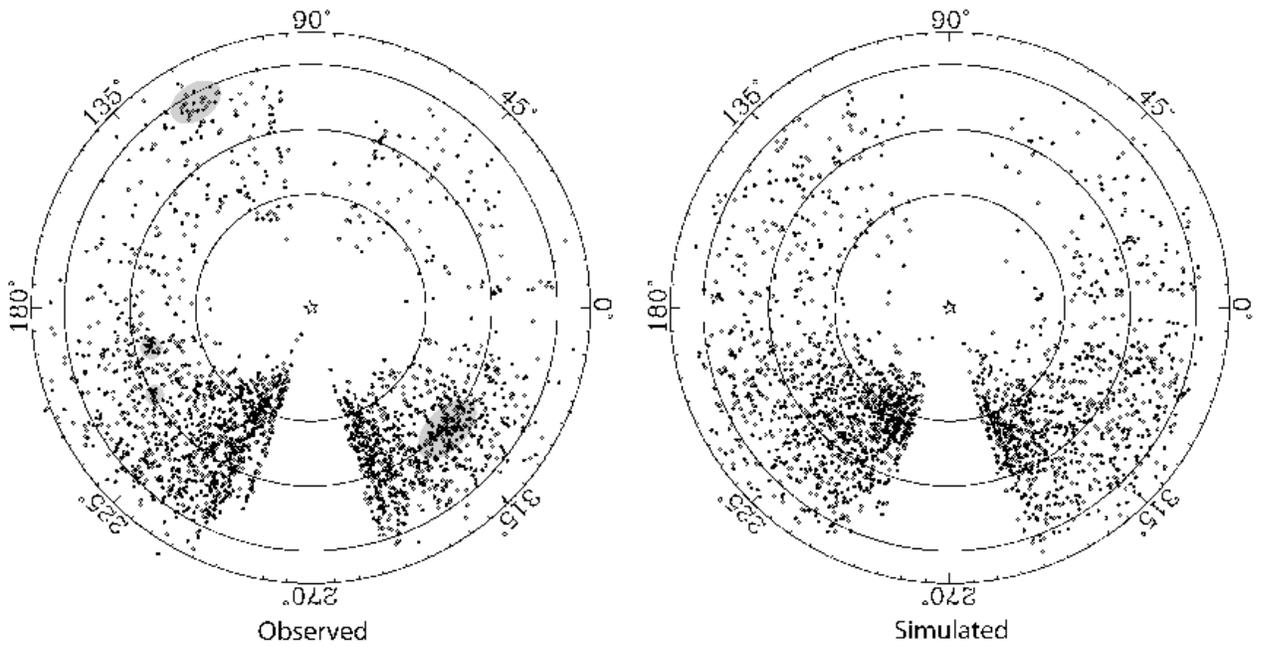}
\caption{\emph{Left:} The radial distribution of RRL candidates in ecliptic longitude. \emph{Right:} The corresponding distribution of objects from one of our monte-carlo simulations of the halo (see \S\ref{section:substructure}). A number of substructures (shaded in the figure) are apparent in the observations that are not found in the simulation, in particular at ecliptic latitude 120$^{\circ}$, 190$^{\circ}$, 210$^{\circ}$, and 320$^{\circ}$. The significance of these structures is discussed in \S\ref{section:substructure}. The circles are placed at $V$=15, 17 and 19.}\label{figure:radial}
\end{center}
\end{figure*}

In Figure \ref{figure:radial} we present the radial distribution of RRL candidates. We collapse the distribution in ecliptic latitude, the natural system for our sample. A series of over densities can be seen, in particular, at ecliptic latitude 120$^{\circ}$, 190$^{\circ}$ and 320$^{\circ}$. These features are however, overshadowed by the spatial non-uniformity of the survey and the general density falloff away from the Galactic centre. In the next two sections we characterise the general candidate background in order to subtract it and reveal the significance of these features.

Figure \ref{figure:SDSSview} shows the position of these over densities on the plane of the sky in the context of known significant substructure. We now discuss these features in more detail:

\subsection{Spatial Density of RRL Candidates}

It is clear from the substructure evident in Figure \ref{figure:radial} that the halo does not exhibit a smooth radial density profile. In order to characterise the significance of the observed substructure we determined the general halo background upon which the substructures are overlaid. 

Note that our sample includes significant contamination by non-RRL objects. It is unlikely that the contaminants share the spatial distribution or absolute magnitude of the RRL population. In the analysis that follows we treat each object as an RRL. The effect this has on the derived spatial density of candidates depends on the nature of the contaminants. For example, main-sequence eclipsing binaries of significantly lower absolute magnitudes than RRLs will appear over a range of magnitudes out to the edge of the disk along a line of sight. Assuming RRL absolute magnitudes for these objects will act to enhance the space density at large distances. The goal of this exercise is not however to derive the spatial density of RRLs but rather to derive a model for the background halo.

There are two key models discussed in the literature for the radial distribution of matter in the halo. General ellipsoidal isopycnic contours for the halo can be expressed as:
\begin{equation}\label{eqn:spherical}
\frac{x^{2}}{a^{2}} + \frac{y^{2}}{b^{2}} + \frac{z^{2}}{c^{2}} = 1
\end{equation}
The spherical model ($a=b=c$) is the most basic, in which the density varies as a function of galactocentric distance as a power law,
\begin{equation}\label{eqn:spherical2}
\rho(R) = \rho_{0}(R/R_{0})^{n},
\end{equation}
where $\rho_{0}$ is the local space density of RRLs and $R_{0}$ is the Solar radius.

The case of a spheroidal halo has $a=b$ and flattening described by $q \equiv c/a$.  Numerous observational studies have found that the inner $<20$kpc of the halo is significantly flattened while the outer regions appear spherical \citep{Chiba00,Preston91}. The spatially variable flattening ratio introduced by \citet{Preston91} expresses isopycnic contours as ellipsoids of revolution. Such surfaces vary as a function of the semi-major axis $a$,
\begin{equation}
c/a = \left\{ \begin{array}{ll}
	(c/a)_{0} + [1-(c/a)_{0}](a/a_{u}), & a<a_{u},\\
	1, & a>a_{u},\\
	\end{array}
	\right.
\end{equation}
where $a_{u} = 20$kpc and canonically, $(c/a)_{0} = 0.5$. In this case, density varies as a function of semi-major axis as:
\begin{equation}\label{eqn:variableflat}
\rho(a) = \rho_{0}(a/R_{0})^{n},
\end{equation}

Numerous studies of the RRL distribution have found that the observed distribution is best fit by variable flattening model. However, the difference in the quality of the resultant fit between the spherical and variably flattened models is of little \citep{VivasZinn06} or no \citep{Miceli07} statistical significance. In the following we only consider the case of a spherical halo.

The calculation of the RRL space density follows the technique developed by \citet{Wetterer96}. The following equation describes the density as a function of galactocentric distance:
\begin{equation}\label{equ:fofr}
\rho(R) = \frac{1}{4 \pi R^{2} f(R)} \frac{dN}{dR},
\end{equation}
where $f(R)$ is the fraction of the total halo volume at $R$ that is sampled by the survey ($f(R)$ is analogous to a solid angle) and N is the number of RRL as a function of distance. Whereas the solid angle is constant as one looks through the halo, $f(R)$ varies as a function of Galactocentric radius and hence must be calculated numerically for each field. This is because the volume of each field is a pyramid subtending the field's effective area (see \S6) at infinity and an apex offset from the Galactic centre by $R_{0}$. Different lines of sight sample different regions of the Galactocentric sphere. This is demonstrated for two lines of sight in Figure \ref{figure:fofr}. One field is of higher galactic latitude and samples more of the inner galactic halo. To generate these profiles we calculate $R$ and the volume of each slice of the pyramid for a given $r+dr$. The volumes are summed into 0.1kpc bins before being divided by the volume of the sphere at the radius $R$ of the bin. As $R$ tends to large values $f(R)$ approaches the asymptotic limit of (effective area)$/4\pi$, corresponding to the constant solid angle as would occur if $R_{0}=0$.

To account for the effects of completeness we mulitply each $f(R)$ by the Monte-Carlo derived RRL$ab$ completeness profile for the field as a function of galactocentric radius. This gives the \emph{effective} volume of the halo that is sampled by each field (the right-hand plot of Figure \ref{figure:fofr}). The total volume of our sample is then simply the sum of $f(R)$ for the 3692 individual fields. This is shown in Figure \ref{figure:fofrtotal}.

We then use Equation \ref{equ:fofr} to calculate the local space density for each candidate. To distill the space density of our sample we then bin space densities of the 2016 RRL candidates in bins of 0.02 dex in log($R$). The resulting average density distribution is shown in Figure \ref{figure:density}. The error bars in Figure \ref{figure:density} represent the standard deviation of the points in each bin. Since 50 to 30 objects reside in a typical bin the large standard deviation is indicative of intrinsic azimuthal variability in the space density such as would result from substructure. 

As stated above, our goal in the examination of the average space density is to extract the mean background upon which the substructure lies. A linear least squares fit to points between $10<R<45$kpc yields $n = -2.48\pm0.09$ and $\rho_{0} = (6.22\pm0.12$)kpc$^{-3}$ (as introduced in Equation \ref{eqn:spherical}). Beyond R=45kpc there is a more rapid drop in the number of candidate RRLs. For the 122 candidates with $R > 45$kpc we find a $n = -5.0\pm0.3$ power-law. We plan to follow-up R$>$45kpc candidates to see if the RRLs amongst them match this increased radial drop-off. However, as stated before, since the candidates contain significant contamination by non-RRLs, we do not interpret these results further at present but rather use them as the basis of our detection of spatial departures from the mean background.



\subsection{Halo Substructure}\label{section:substructure}

In this section we quantify the significance of substructure in the halo. We firstly construct a series of artificial halos based upon the spatial distribution function discussed above. This is complicated in the case of our survey by non-uniform spatial, temporal and radial coverage, however all three are well characterised for each field. To form a simulated halo we first randomly select a position in the sky from the area covered by our survey fields. We then use the spatial density power-law indices discussed above to randomly assign a distance and then use the completeness information for the field in which it lies to see whether we would recover such a candidate. This is repeated until the total number of candidates observed is recovered. An example of a simulated halo is shown in Figure \ref{figure:radial}.

We next perform a Voronoi tessellation \citep{Voronoi, Ramella01} of the simulated halo distribution in (heliocentric distance, ecliptic longitude) space. The Voronoi tessellation associates with each point a polygon such that every point inside the polygon is closest to the point in question. It enables us to define the local density of points in a non-parametric way as simply the reciprocal of the area of the tesserae. We repeat this for the entire set of simulated halos and form a map of the mean density of simulated candidates in (distance, ecliptic longitude) space. Along with the mean we determine a map of the standard deviation in the simulated density. From the Voronoi tessellation of the observed RRL candidate distribution we subtract the mean density of simulated candidates and divide by the standard deviation to obtain a map of the significance of spatial structure in the halo as seen in Figure \ref{figure:80kpc_Sgr_view}.

\subsubsection{Sagittarius Dwarf Debris Streams}

\begin{figure*}
\begin{center}
\includegraphics[scale=0.7, angle=0]{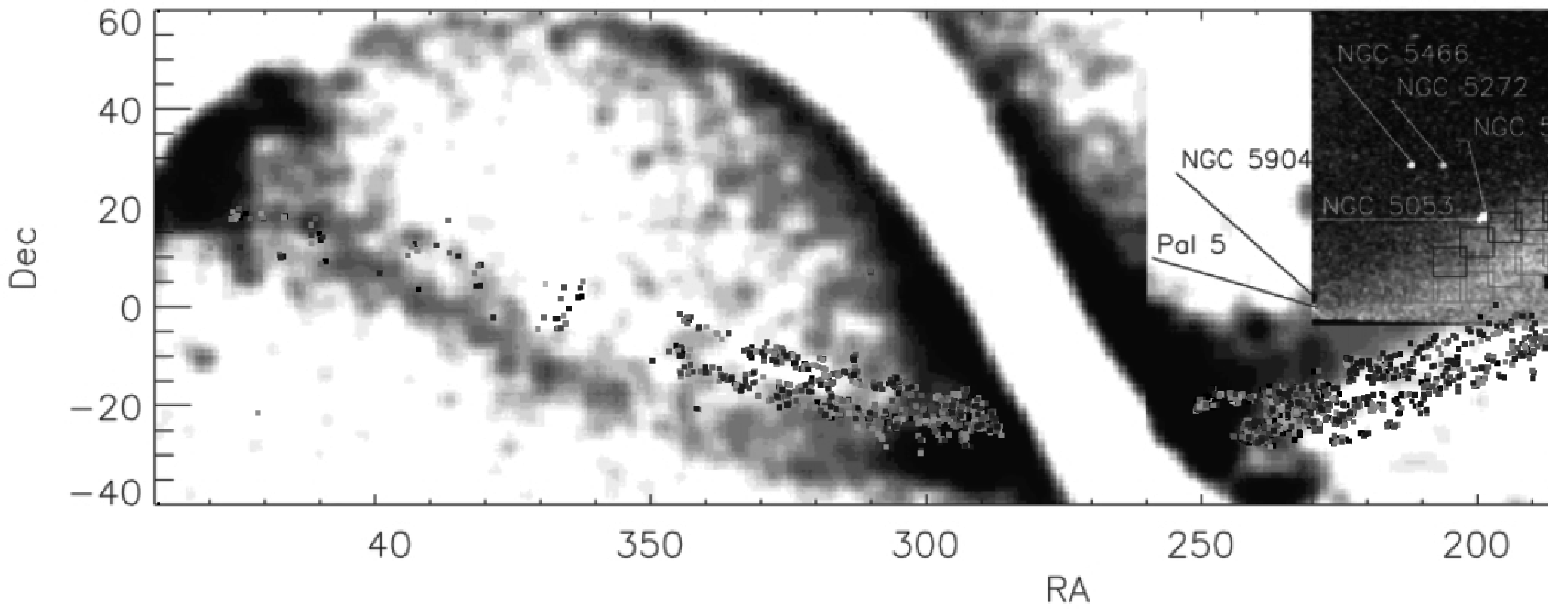}
\caption{Our 2016 RRL candidates color coded by distance overlaid upon the view of the outer halo due to \citet{Belokurov07} which includes analysis of the main-sequence turnoff stars from SDSS (the top right insert) and M giants from the study of \citet{Majewski03} (the greyscale background). The Sagittatius Stream dominates the view of the outer halo (the main body of the Sagittarius dwarf is located centrally in this figure at Dec$\sim-20^{\circ}$. See the electronic edition of the Journal for a color version of this figure.}\label{figure:SDSSview}
\end{center}
\end{figure*}

\begin{figure}
\begin{center}
\includegraphics[scale=0.38, angle=0]{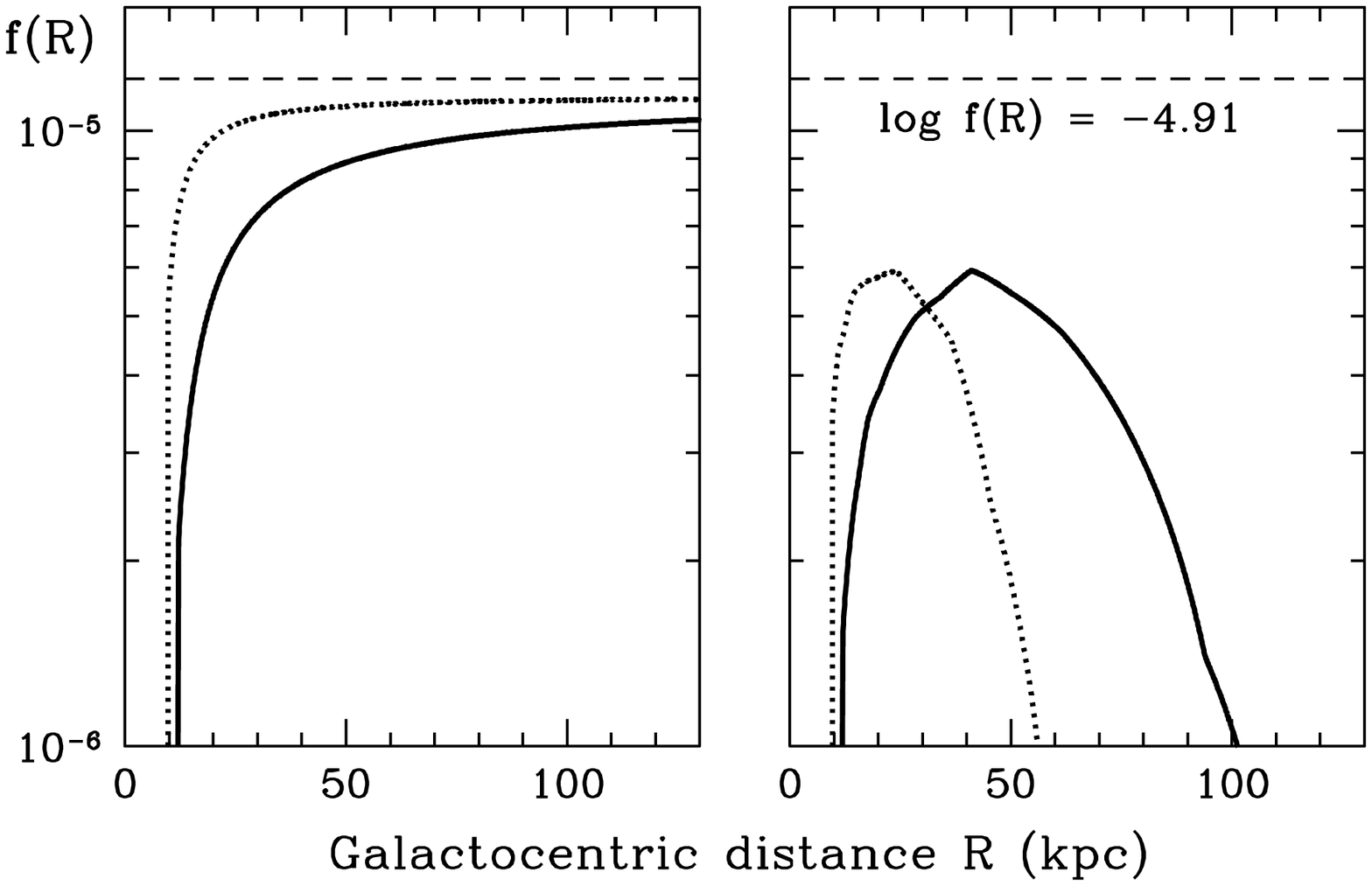}
\caption{Variation of the fractional halo volume as a function of Galactocentric distance, $f(R)$. \emph{Left panel:} for two fields centred on $(\ell,b)=(213^{\circ},34^{\circ})$ (solid line) and $(262^{\circ},60^{\circ})$ (dotted line). The asymptotic limit represents the constant solid angle subtended by the fields if $R_{0}=0$ with infinite limiting magnitude. \emph{Right panel}: The same profiles corrected for completeness as a function of $R$. The profiles for all fields are summed to give the total $f(R)$ for the survey. }\label{figure:fofr}
\end{center}
\end{figure}

\begin{figure}
\begin{center}
\includegraphics[scale=0.38, angle=0]{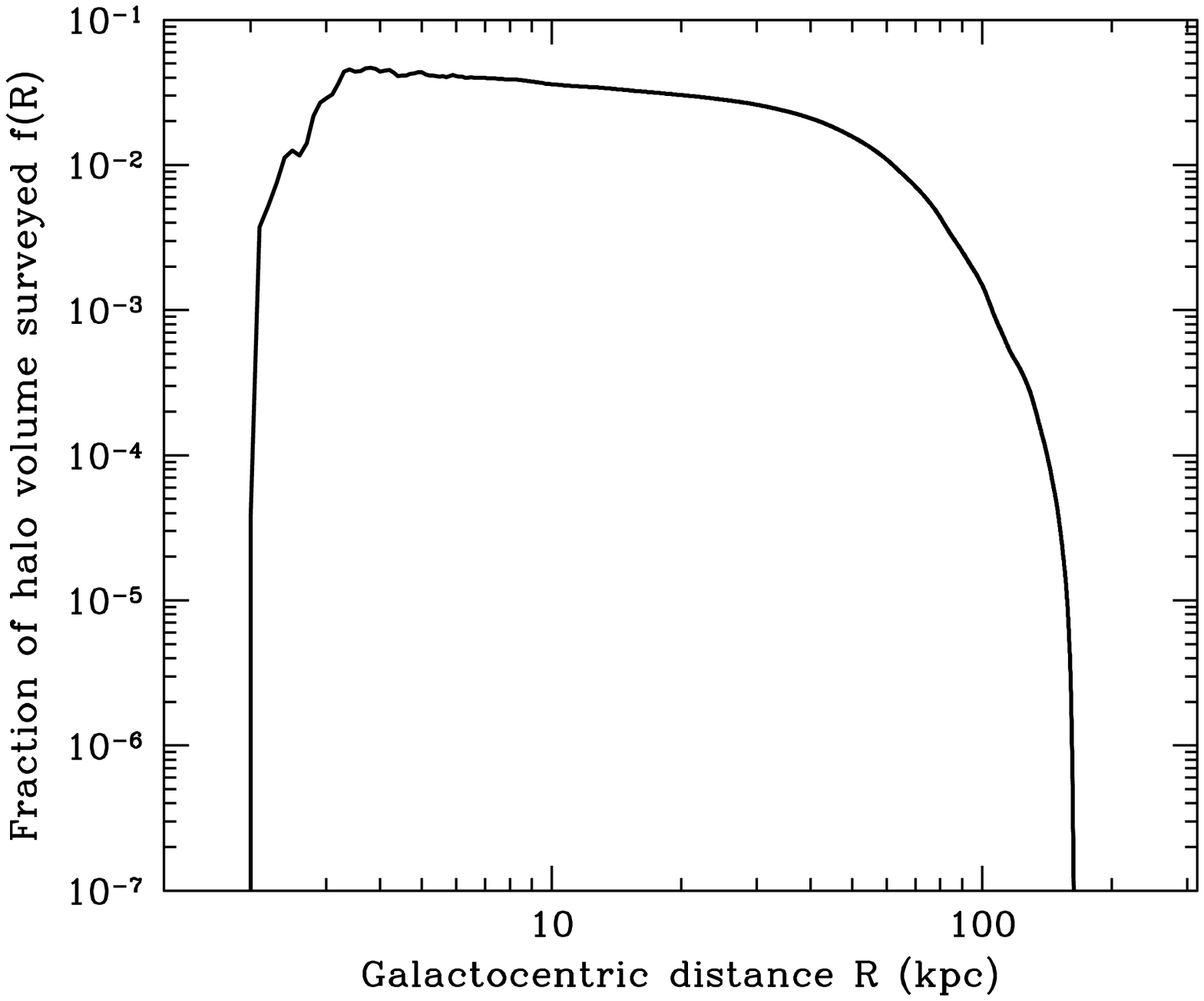}
\caption{Total fractional volume of the halo surveyed by the 3692 fields}\label{figure:fofrtotal}
\end{center}
\end{figure}

\begin{figure}
\begin{center}
\includegraphics[scale=0.38, angle=0]{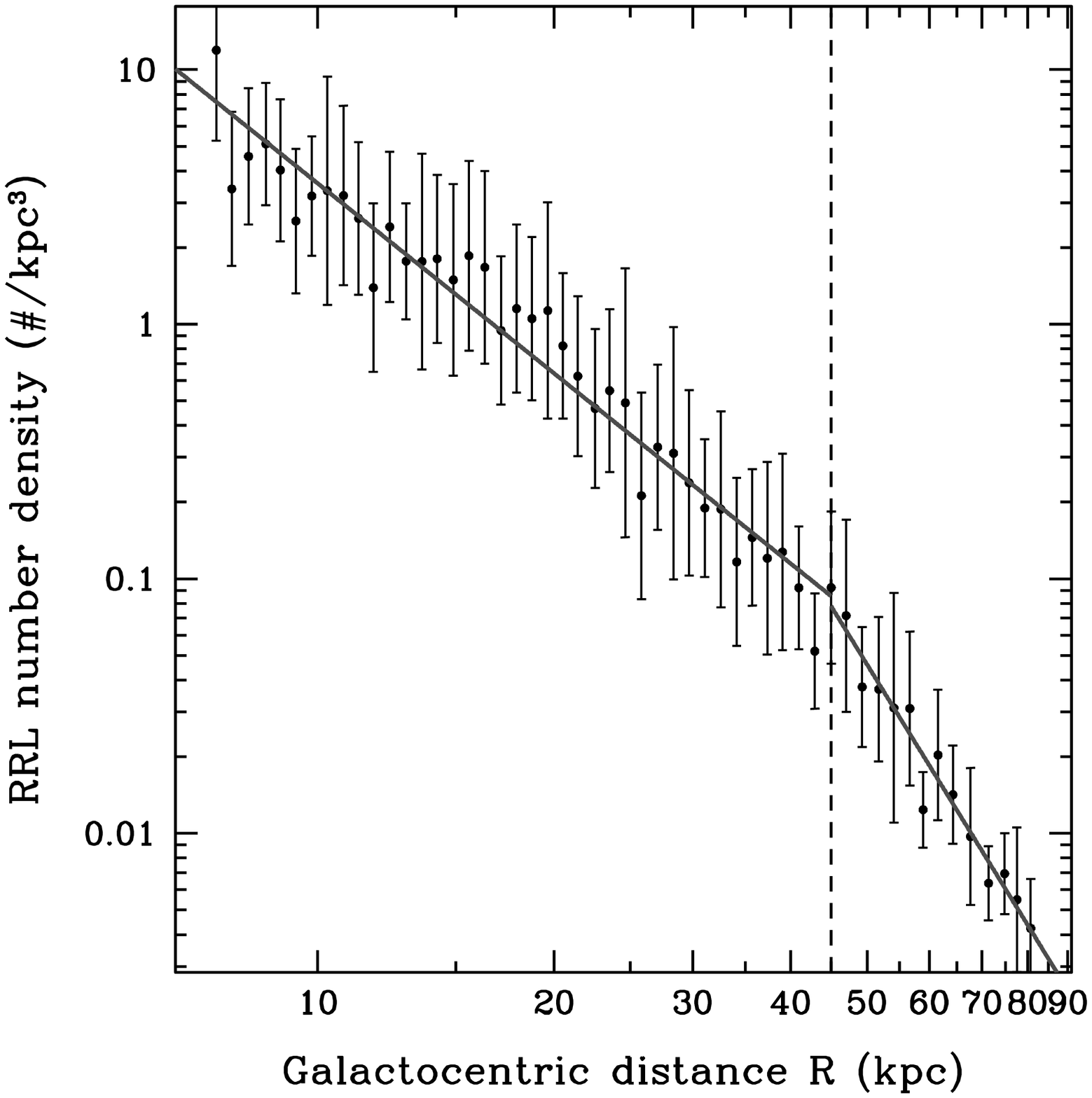}
\caption{The apparent spatial density of RRL candidates determined assuming that all candidates are RRLs (see text for discussion of the best-fit lines). We use this result to quantify the average background upon which substructure resides.}\label{figure:density}
\end{center}
\end{figure}

\begin{figure*}
\begin{center}
\includegraphics[scale=0.70, angle=0]{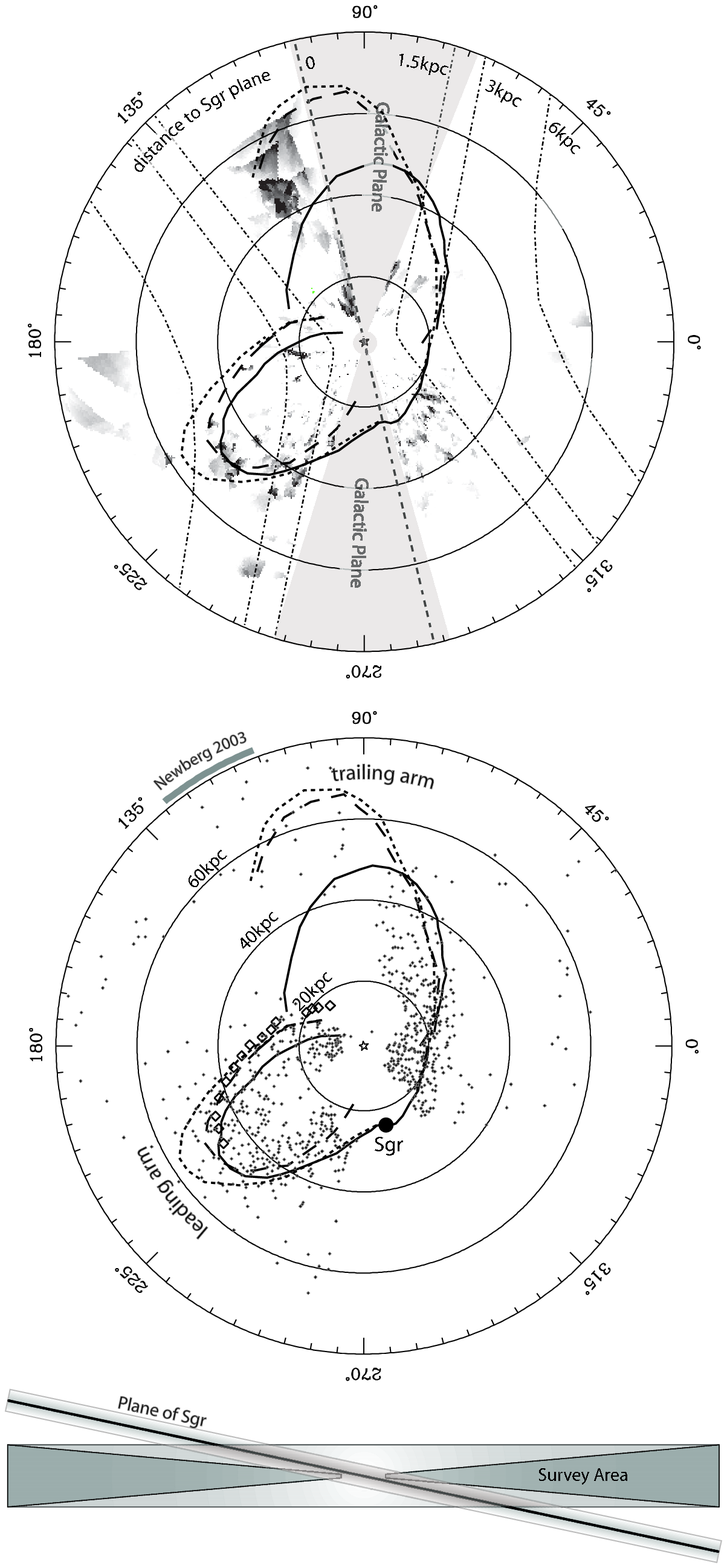}
\caption{\emph{Top:} A significance map of over densities in RRLs in the halo (see text for details on derivation). Regions in black represent over densities of 8$\sigma$ significance. The circles are at heliocentric distances of 20, 40, 60 and 80kpc. The dashed straight line is the line of nodes for the intersection between the orbital plane of the Sagittarius dwarf galaxy \citep{Ibata03} and the ecliptic along which our survey is performed. The dotted lines are distances at which the plane of the Sagittarius dwarf is 1.5, 3, and 6kpc away from the survey volume. \emph{Middle:} A view of the Sagittarius plane showing M giants from \citet[points]{Majewski03}, \citet[squares]{fieldofstreams} distances and detections of Sgr stars from \citet[RA$\sim120^{\circ}$; 83kpc]{Newberg03}. Shown are models from \citet[solid line]{Law05}, \citet[dashed]{Fellhauer06}, and \citet[short dashed]{Helmi01}. \emph{Bottom:} a schematic of the geometry of Sgr relative to our survey volume as seen along the line of nodes. }\label{figure:80kpc_Sgr_view}
\end{center}
\end{figure*}

In the interpretation of Figure \ref{figure:80kpc_Sgr_view} it is useful to consider the position of the orbital plane of the Sagittarius dwarf galaxy that features prominently in the outer halo (as seen in Figure \ref{figure:SDSSview}). Our survey cuts obliquely through the plane of the Sagittarius dwarf (Sgr; as defined by \cite{Ibata01}) at an angle of 16$^{\circ}$ with an opening angle of 10$^{\circ}$ (see Figure \ref{figure:80kpc_Sgr_view}). We obtain maximum signal from structures in the Sagittarius Stream at angles adjacent to the line of nodes. The structures from RA=45$^{\circ}$ to 130$^{\circ}$ and 220$^{\circ}$ to 330$^{\circ}$ lie in this range. In Figure \ref{figure:80kpc_Sgr_view} we show a view of the Sagittarius tidal arms compiled from the observations of \citet[M giants]{Majewski03}, \citet[F dwarfs]{fieldofstreams} and \citet[Blue horizontal branch stars and F dwarfs]{Newberg03,Newberg07} together with the model predictions of \citet{Helmi01}, \citet{Law05}, and \citet{Fellhauer06}.

The models of \citet{Law05} and \citet{Helmi04} aim to match the observations of M giants \citep{Majewski03}. \citet{Helmi04} finds that the data strongly suggests that the dark matter halo of the Milky Way is prolate (($1.25 < (q = a/c) < 1.5$). \citet{Law05} similarly require the presence of a significantly prolate halo ($q = 1.25$) but could not establish a model that reproduces both the observed orbital pole precession and the distance and radial velocity of the leading arm. \citet{Johnston05} conclude that it is necessary to consider an evolution in the  orbital parameters of Sgr over the timeframe of the last few orbits ($\sim$2-3 Gyr) and propose that dynamical friction is the most favourable mechanism to bring about this change. 

\citet{Fellhauer06} models the Sgr system in the light of the \citet{fieldofstreams} result. In particular, Fellhauer et al.\ seeks to interpret the observed bifurcation of the Sgr stream. This work is based upon the distances derived to the F dwarf population. As pointed out by \citet{Newberg03} a 13\% discrepancy in distance exists between the distances derived from M giants and those from F dwarfs. The M giant distance scale has proven to be systematically smaller than the F dwarf scale and it is unclear which is more accurate. Fellhauer et al.\ finds that the observed bifurcation is only possible in a near-spherical halo and that it is not possible to match the distances and radial velocities of the leading arm in their closest approach to the Solar circle. We must bear in mind these caveats to the modelling of the debris from Sgr as we now discuss the over densities found in the present study with reference to the orbit of Sgr.

The Sgr leading arm is potentially seen in the over density located at RA=115$^{\circ}$ and distance of 15kpc (see Figure \ref{figure:80kpc_Sgr_view}). This region is a confluence of the Monoceros Ring and Sagittarius Stream both located at 15-20kpc in this region \citep[respectively]{Ibata03,fieldofstreams}. The Sgr leading arm then passes above our survey area until it is again intercepted at RA=220$^{\circ}$, d=50kpc. This region is the southern limb of the Sgr leading arm (see Figure \ref{figure:SDSSview}) the northern edge of this structure is seen in the QUEST survey \citep[their Figure 12]{VivasZinn06}. The main body of Sgr lies in the region of the Galactic plane avoided by our survey, however, we see over densities coincident with the trailing arm from RA=287$^{\circ}$ to 330$^{\circ}$, d=25kpc. The trailing arm then passes underneath our survey area before potentially emerging at RA=130$^{\circ}$, d=45kpc.

The over-density at (RA=130$^{\circ}$, d=45kpc) is particularly interesting. It is separated by $\sim200^{\circ}$ from the main body of Sgr along the trailing arm. Other than the over-density of A-type stars detected by \citet{Newberg03}, this over-density is the most distant and oldest portion of Sgr debris yet found and offers a critical constraint on the orbit of Sgr and hence the potential of the Galaxy in which it moves. As we can see from the divergence between the models of \citet{Law05} and \citet{Fellhauer06} in this region (see Figure \ref{figure:80kpc_Sgr_view} middle panel), older material offers the strongest constraint on the model orbit. However, radial velocities for stars in the observed over-densities are required to confirm their association with Sgr. We are now obtaining radial velocities for RR Lyraes found along the debris stream. By confronting the radial velocity predictions of dynamical models with observations we hope to converge on increasingly accurate models for the Galactic halo and orbital elements for Sgr.

\subsubsection{Virgo Over Density}

\begin{figure}
\begin{center}
\includegraphics[scale=1.20, angle=0]{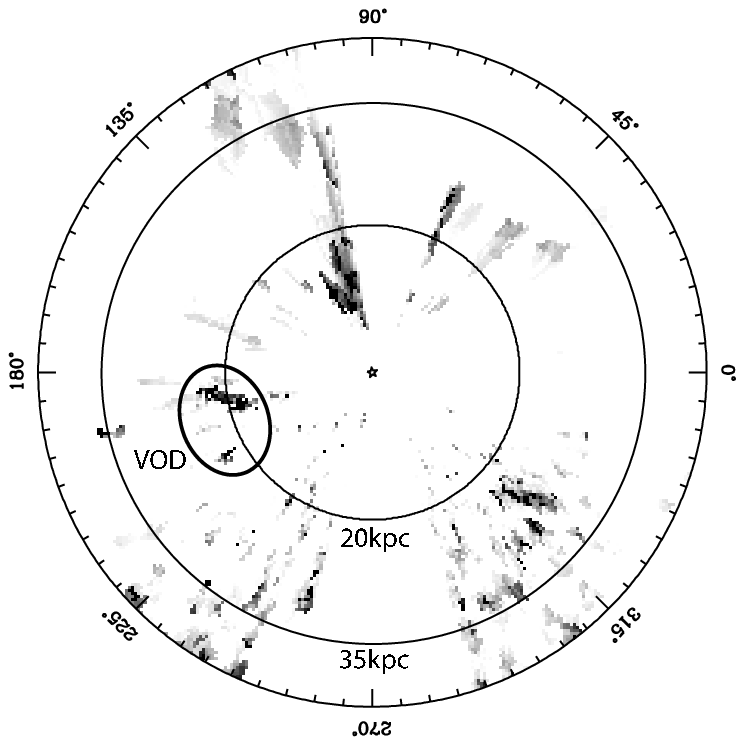}
\caption{Close-up view of the inner 40kpc of our survey. While much of the structure can be attributed to the debris of Sgr, the Virgo Over Density region at RA$\sim180^{\circ}$-210$^{\circ}$ and distance $\sim20$kpc is evident. Circles are at heliocentric distances of 20 and 35kpc and angles are in degrees of right ascension.}\label{figure:VOD}
\end{center}
\end{figure}

The region between RA 180-230$^{\circ}$ (seen in Figure \ref{figure:VOD}) exhibits two significant over densities, one at RA=12$^{h}$47$^{m}$, dec=-06$^{\circ}$42$^{m}$ and distance of 16kpc and a second at RA=13$^{h}$47$^{m}$, dec=-11$^{\circ}$24$^{m}$ and a distance of 19kpc. Both lie in the direction of several  previously identified halo substructures. The first indication of substructure in this region was presented by \citet[region S297+63-20.5; at a distance of 18kpc]{Newberg02}. \citet{Zinn04} found a significant over  density from their RRL survey (distance of 19kpc). Subsequent spectroscopy of a sample of RRLs and blue horizontal branch stars from within this clump revealed a moving group with $V_{gsr}=100\pm13$kms$^{-1}$ \citep[dubbed the Virgo Stellar Stream]{Duffau06}. \citet{Juric05} used photometric parallaxes of SDSS stars and found a broad over density centred on the above (the Virgo Over Density (VOD)) that covers over a thousand square degrees between distances of 5-15kpc. \citet{MartinezDelgado07} suggests that the leading arm of tidal debris from the Sagittarius dwarf, as modelled by \citet{Law05} could provide the excess required to account for the VOD. \citet{NewbergYanny05} have suggested the VOD may result from a non-axisymetric component to the Galaxy such as a triaxial halo. 

\citet{Newberg07} determines radial velocities for a sample of F stars drawn from the S297+63-20.5 region and finds radial velocities of $V_{gsr}=130\pm10$kms$^{-1}$. The low velocity dispersion of these stars is inconsistent with their origin in a triaxial halo and their radial velocity is at odds with that expected from stars in the Sgr leading tidal tail. \citet{Newberg07} proposes that the three named features are one, we use the term Virgo Over Density hereafter to refer to this single system.

\begin{figure}
\begin{center}
\includegraphics[scale=0.75, angle=0]{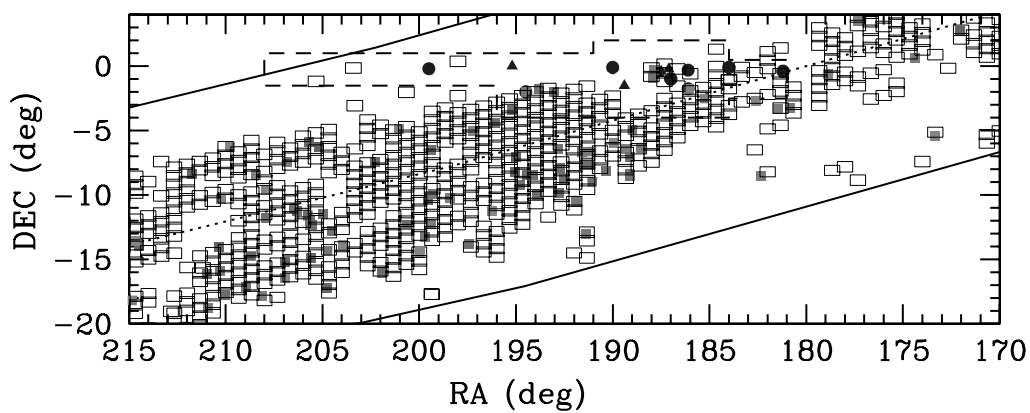}
\caption{The distribution of RRL candidates in the region of the Virgo Over Density covered by our survey (survey fields are the open squares). The points shown have heliocentric distances of $16 < r < 26$kpc. Clump 1 is seen at RA=192$^{\circ}$, dec=-8$^{\circ}$ and Clump 2 at RA=206$^{\circ}$, dec=-12$^{\circ}$. The sample of RRLs and blue horizontal branch stars of \citet{Duffau06} are shown as filled circles and triangles respectively. The dashed outline shows the region where Duffau et al.\ found an excess of main-sequence stars of appropriate apparent magnitude to be associated with the Virgo Over Density. }\label{figure:VODspatial}
\end{center}
\end{figure}

The first over density, VOD Clump 1, lies 8 degrees southwest of the region identified by \citet{Duffau06} as shown in Figure \ref{figure:VODspatial}. This is followed by a second over density 16$^{\circ}$ further to the south west, VOD Clump 2. It is apparent from the results of \citet{fieldofstreams} and \citet{Newberg07} that regardless of the likely shape of the VOD the central regions must lie out of the SDSS region in the southern hemisphere. The twin clumps found by the present study plausibly form structure in the core of the VOD. However we must point out that due to the non-uniform survey coverage we are unable to clearly define spatial substructure. We are in the process of attaining radial velocities for samples from the two clumps to verify if they do indeed share the systemic velocity found by \citet{Duffau06} for the VOD.

\section{Summary}

In this paper we have presented a survey for RR Lyrae stars that has revealed a sample of 2016 RR Lyrae candidates from multiple (2-10) epoch data covering 1675 square degrees. Follow up photometric observations of these candidates show 76\% to be RR Lyrae variables. Our survey reveals a series of significant over densities in the Galactic halo. Many of the over dense regions lie along the leading and trailing arms of debris from the Sagittarius dwarf galaxy extending up to $\sim200^{\circ}$ from the main body of the Sagittarius dwarf. Another locality of over density can not be ascribed to the Sagittarius dwarf: namely the region of the Virgo Over Density. Here we find two significant over densities 10$^{\circ}$ to the south west of the centre previously reported. 

Radial velocities of RR Lyrae stars drawn from our over densities will give insight into the dynamics and origins of these systems. In particular, regions coincident with the leading and trailing arms of the Sagittarius dwarf could offer powerful constraints on the orbit and evolution of the Sagittarius dwarf.



\acknowledgments

We thank Ken Freeman for his comments that helped to improve this paper. This data is based upon observations from Mount Stromlo Observatory's 1.3m Telescope before its untimely demise in the bushfire of January 18th 2003. This paper utilises public domain data obtained by the MACHO Project, jointly funded by the US Department of Energy through the University of California, Lawrence Livermore National Laboratory under contract No. W-7405-Eng-48, by the National Science Foundation through the Center for Particle Astrophysics of the University of California under cooperative agreement AST-8809616, and by the Mount Stromlo and Siding Spring Observatory, part of the Australian National University. This publication makes use of data products from the Two Micron All Sky Survey, which is a joint project of the University of Massachusetts and the Infrared Processing and Analysis Center/California Institute of Technology, funded by the National Aeronautics and Space Administration and the National Science Foundation.



{\it Facilities:} \facility{MtS:1.3m}.

\clearpage



\clearpage

\clearpage

\clearpage

\clearpage

\clearpage

\clearpage

\clearpage

\clearpage

\clearpage

\clearpage

\clearpage

\clearpage

\clearpage

\clearpage

\clearpage

\clearpage

\clearpage

\clearpage

\clearpage

\clearpage

\clearpage

\clearpage

\begin{table}
\caption{Wide-field Surveys for RRLs.}
\begin{center}
\begin{tabular}{cccc}
\tableline\tableline
 Survey &  Sky Coverage &  Number of    &  Maximum \\
 Name  &     (sq. deg.)       &   Candidates &  Distance(kpc)  \\
\tableline
 QUEST &  380 &  498 &  55\\
 LONEOS-1 &  1430 &  838 &  30\\
 Ivezic et al.\ (2000) &  100 &  296 &  65\\
 Sesar et al.\ (2007)& 290 &  634 &  100\\
 Current Survey &  1675 &  2016 &  50\\
\tableline
\end{tabular}
\end{center}
\label{table:surveys}
\end{table}

\clearpage

\begin{table}
\begin{center}
\caption{Table of RR Lyrae candidates found in the present work. }
\begin{tabular}{lcccccc}
\tableline\tableline
 Object ID &  RA &  Dec &  V &  V-R &  E(V-R) &  Heliocentric Distance (kpc)\\
\tableline
128416.544 & 00:12:52.83 & 05:02:56.71 & 17.17 & 0.19 & 0.014 & 20.3 \\
116375.115 & 00:13:28.83 & 02:03:43.68 & 14.50 & 0.08 & 0.015 & 5.9 \\
128412.1045 & 00:14:37.09 & 03:51:27.99 & 18.35 & 0.20 & 0.013 & 35.0 \\
115608.149 & 00:16:14.73 & 01:53:51.58 & 15.34 & 0.18 & 0.019 & 8.6 \\
\tableline
\end{tabular}
\label{table:data}
\end{center}
\tablecomments{Table \ref{table:data} is published in its entirety in the 
electronic edition of the {\it Astrophysical Journal}.  A portion is 
shown here for guidance regarding its form and content.}
\end{table}






\end{document}